\documentclass[proof]{WileyASNA-v1}

\articletype{Article Type}%

\received{26 April 2016}
\revised{6 June 2016}
\accepted{6 June 2016}

\usepackage{graphicx,times}             %for PS/EPS graphics inclusion, new
\usepackage{natbib}
\usepackage{amssymb,amsmath}
\usepackage{upgreek}
\usepackage{nicefrac}
\usepackage{booktabs}
\usepackage[flushleft]{threeparttable}
\usepackage{siunitx}
\usepackage{comment}
\usepackage{CJK} 

\newcommand{\kepler}{Kepler}
\newcommand{\tess}{TESS}
\newcommand{\corot}{CoRoT}
\newcommand{\numax}{\mbox{$\nu_{\rm max}$}}

\newcommand{\muHz}{\mbox{$\mu$Hz}}

\newcommand{\msun}{\!{\mathrm M_{\odot}}}

\begin{document}
%\begin{CJK*}{UTF8}{gbsn}

\title{Contributions of Structural Variations to the Asymptotic Mixed-mode Coupling Factor in Red Giant Stars}

\author[1]{Chen Jiang*}

\authormark{CHEN JIANG}

\address[1]{\orgname{Max-Planck-Institut f\"ur Sonnensystemforschung}, \orgaddress{\state{Justus-von-Liebig-Weg 3, 37077 G\"ottingen}, \country{Germany}}}

\corres{*Chen Jiang, Max-Planck-Institut f\"ur Sonnensystemforschung, Justus-von-Liebig-Weg 3, 37077 G\"ottingen, Germany. \email{jiangc@mps.mpg.de}}

\presentaddress{Max-Planck-Institut f\"ur Sonnensystemforschung, Justus-von-Liebig-Weg 3, 37077 G\"ottingen, Germany}

\abstract{The advent of ultra-precise photometry space missions enable the possibility of investigating stellar interior with mixed modes. The structural variations induced by the discontinuity of the chemical composition left behind during the first dredge--up is an important feature in the stellar mid-layers located between the hydrogen-burning shell and the base of the convective zone of red giants, as the mixed-mode properties can be significantly affected by these variations. In this paper, the contributing factors to variations of the mixed-mode coupling factor, $q$, are discussed with stellar models. In general, the structural variations give rise to a subtle displacement in the Lamb frequency and a sharp change in the buoyancy frequency, which lead to variations in the value of $q$ computed using the asymptotic formalisms that assuming a smooth background free of structural variations. The impact of these two factors can be felt in detectable mixed modes in low-luminosity red giants. Furthermore, the different nature of variations of the two characteristic frequencies with radius near the base of the convective zone,  produces a sudden increase in $q$ in evolved red giants. This is followed by a quick drop in $q$ as the star evolves further along the red giant branch. }


\jnlcitation{\cname{%
\author{Jiang C.}} (\cyear{2022}), 
\ctitle{Contributions of Structural Variations to the Asymptotic Mixed-mode Coupling Factor in Red Giant Stars}, \cjournal{Astronomische Nachrichten}, \cvol{2022;00:1--6}.}

%%\fundingInfo{Funding info text.}

\maketitle

\section{Introduction} 
\label{sect:intro}

The development of space-born missions dedicated to observe stellar oscillations, such as \corot\ \citep[e.g.,][]{baglin2006}, \kepler\ \citep[e.g.,][]{borucki2010} and \tess\ \citep[e.g.,][]{ricker2014}, allow access to abundant and ultra-precise seismic data to study the stellar inner structure for giant stars \citep[e.g.][for a review]{hekker2017}. One of the important advances in this context is the investigation of internal gravity modes and mixed modes in these stars \citep[e.g.,][]{mosser2016}. Waves of mixed modes have simultaneously the properties of an acoustic wave in the stellar envelope and the properties of a gravity wave in the inner radiative regions near the core. The mixed pressure-gravity nature allows for detection in subgiant and red-giant stars, in which the inner core condenses dramatically leading to an increase in the local gravitational acceleration, and hence in the buoyancy frequency. This gives rise to the mixing between pressure and gravity waves. Mixed modes in subgiants and red giants have been widely analysed with \corot\ and \kepler\ data \citep[e.g.,][]{carrier2005, bedding2007, huber2010, jiang2011, kallinger2012}. 

Studies of the gravity-mode property of mixed modes enabled the possibility to probe the structure of the core by estimation of the g-mode period spacing of mixed modes \citep[e.g.][]{bedding2011}, and allowed monitoring the stellar interior rotation through identification of frequency splittings \citep[e.g.][]{beck2012, gizon2013}. In addition, analyses of other seismic parameters associated with mixed modes, such as the gravity offset \citep{pincon2019} and the coupling factor \citep{pincon2020}, helped to reveal information of the structure of the mid-layers near the base of the convective zone (BCZ) in red giants. The coupling factor $q$ is the dimensionless measurement of the mixture of the pressure mode (p-mode) and gravity mode (g-mode). According to asymptotic analyses of mixed modes \citep[e.g.,][]{shibahashi1979, takata2016}, the value of $q$ is closely related to the width of the evanescent zone (EVZ) located between the p and g mode cavities and near the stellar mid-layers. Thus, understanding the mid-layer structure of red giants is of great importance to the measurement of mix-mode properties.

An interesting feature of the mid-layer structure is the structural variations due to the contraction and expansion of the convective envelope in post-main-sequence stars. On the subgiant and early red-giant branch (RGB), the outer convective envelope deepens in mass as the star evolves in the so-called \textit{first dredge-up} phase, reaching regions in which chemical composition has been modified by nuclear reactions. This brings about a discontinuity in the hydrogen profile, producing a sharp variation in the density gradient and thus the buoyancy frequency at the BCZ in models without proper treatment of diffusion. As the star evolves further on the RGB, the convective envelope retracts in mass, leaving behind the chemical discontinuity and the sharp variation in the radiative zone, and potentially in the g-mode cavity as the sharp variation of the buoyancy frequency then has the character of a so-called \textit{buoyancy glitch} that can significantly affect the properties of detectable mixed modes, e.g. mode frequencies and inertias \citep{cunha2015, cunha2019}. Consequently, the value of the inferred $q$ of mixed modes is also affected if the impact of the buoyancy glitch is not explicitly accounted for \citep{jiang2020}. These changes in mixed-mode properties due to the buoyancy glitch can be observed in low-mass stars at the red-giant bump \citep{cunha2015} when the stellar luminosity temporarily decreases with evolution as the hydrogen-burning shell approaches the composition discontinuity and starts to feel the lower mean molecular weight above the discontinuity. The luminosity resumes rising as soon as the shell reaches and eliminates the discontinuity \citep{jcd2015} as well as the glitch. 

However, for red giant stars that are much younger than the luminosity bump, the sharp variation in the buoyancy frequency located at the BCZ has a much smaller scale than the glitch, and thus it is called the \textit{buoyancy spike}. In these low-luminosity RGB stars the buoyancy spike is outside the propagation cavity of detectable mixed modes, but can reach the outer edge of the g-mode cavity of low-frequency modes, thus making these modes not free of impact. Moreover, the buoyancy spike being located outside the g-mode cavity means that it is indeed inside the EVZ where the asymptotic $q$ is calculated. Hence the spike inevitably affects the inferred asymptotic $q$. Indeed, \cite{jiang2022} performed asymptotic analyses on mixed modes in low-luminosity RGB models and found that the impact of the buoyancy spike on the mixed-mode properties is significant in models with a large frequency separation in the interval 5 to 15 $\mu$Hz and with an evanescent zone located in a transition region that may be thin or thick.
Since the asymptotic analyses assume a smooth background, they are expected to fail in the presence of the buoyancy spike (glitch) as shown by \cite{jiang2022}.
However, a comprehensive study of the structural variations in terms of understanding how they impact the oscillations and what are the contributions to this impact is still needed and can be done by detailed investigations of the mixed-mode properties in layers near the buoyancy spike (glitch), which is the main focus of this work.

In this paper, the mid-layer structure of RGB stars are analysed using stellar models in an analytical way, to study the impact of the structural variations on dipolar mixed modes (modes with angular degree $\ell = 1$). The paper is segmented into 4 sections. Section~\ref{sc:asyrgb} gives introduction to the asymptotic analysis of mixed modes in RGB stars and how the inferred asymptotic $q$ changes corresponding to the structural variations in the mid-layers. A detailed discussion on three different contributing factors associated with the changes in asymptotic $q$ is presented in Section~\ref{sc:contribution}. Finally, a conclusion is given in Section~\ref{sc:conclusion}.

\section{Asymptotic Coupling factor of mixed modes in red giants}
\label{sc:asyrgb}

\subsection{Asymptotic Formalisms of Coupling Factor}
\label{sc:asyp}

The asymptotic analyses of mixed modes have been extensively discussed in early works \citep[e.g.,][]{shibahashi1979, tassoul1980, unno1989} for stellar nonradial oscillations. Under assumptions of  Cowling approximation \citep{cowling} and a very thick EVZ (normally true for evolved RGB stars) corresponding to very weak mixing between p and g modes, the coupling factor is asymptotically defined by
\begin{equation}
q_{\rm weak} = \frac{1}{4} \exp \left(-2 \int_{\rm EVZ} |K| \mathrm d \,r \right)
\label{eq:qweak}
\end{equation}
with wavenumber $K$ given by
\begin{equation}
K^2 \approx \frac{\omega^2}{c^2} \left( 1 - \frac{S_\ell^2}{\omega^2} \right) \left( 1 - \frac{N^2}{\omega^2} \right),
\label{eq:k}
\end{equation}
where $c$ is the sound speed, $\omega$ is the angular frequency of the oscillation mode, and $N$ and $S_\ell$ are the two characteristic frequencies, the Brunt-V\"ais\"al\"a frequency $N$ (or buoyancy frequency) and the Lamb frequency $S_\ell$ for mode degree $\ell$ \citep[cf. section 3 of][]{aerts2010} that are defined in terms of the density $\rho$, the pressure $p$, the gravitational acceleration $g$ and the first adiabatic index $\Gamma_1$ as
\begin{align}
N^2 &= g \left( \frac{\mathrm d \ln p}{\Gamma_1 \mathrm d r}  - \frac{\mathrm d \ln \rho}{\mathrm d r}\right), \\
S_\ell^2 &= \frac{\ell(\ell+1)c^2}{r^2}.
\end{align}
The integration in equation~\eqref{eq:qweak} is made over the distance encompassing the EVZ, in terms of the radius $r$. Equation~\eqref{eq:qweak} predicts that the asymptotic coupling factor is in the interval 0 to 0.25 in the case of weak coupling in evolved RGB stars.

However, for low-luminosity RGB stars, the EVZ is not very thick and hence equation~\eqref{eq:qweak} is no longer applicable. By fully accounting for the effect of the perturbation of the gravitational potential that is neglected by the Cowling approximation, \citet[][hereafter, T16]{takata2016} introduced an asymptotic formalism for the coupling factor in the limiting case of a very thin EVZ corresponding to strong coupling, which is usually the case observed in subgiants, early red giants and red clump stars \citep{mosser2012}. 
According to T16, the quantities related to the differential equations for dipolar oscillations are defined in terms of the stellar properties as
\begin{align}
\mathcal V &= - \frac{1}{\Gamma_1 \mathcal{J}} \frac{\mathrm{d} \ln p}{\mathrm{d} \ln r},
\label{eq:vr} \\
\mathcal A &= \frac{1}{\mathcal J} \left( \frac{1}{\Gamma_1} \frac{\mathrm{d} \ln p}{ \mathrm{d} \ln r} - \frac{\mathrm{d} \ln \rho}{ \mathrm{d} \ln r} \right),\\
\lambda &= \frac{\omega^2 r }{g},\\
\mathcal J &= 1 - \frac{1}{3} \frac{\mathrm d\, \ln M_r}{\mathrm d\, \ln r} = 1 - \frac{4}{3}\frac{\uppi \rho r^3}{M_r},
\label{eq:j}
\end{align}
with $M_r$ being the concentric mass contained in radius $r$. Since the average density of $M_r$ is larger than the density $\rho$ at radius $r$, except at the centre, $M_r$ is always larger than $\frac{4}{3}\uppi \rho r^3$. As a result, $\mathcal J$ is positive but less than 1. Then we can define
\begin{equation}
P = 2 \mathcal J - \lambda \mathcal V
\label{eq:pr}
\end{equation}
and 
\begin{equation}
Q = \mathcal J - \frac{\mathcal A}{\lambda}.
\label{eq:qr}
\end{equation}
It can be proved that $\sqrt{PQ}/r$ reproduces the radial wavenumber in the p and g mode cavities of \citet[][cf. sections 15 and 16]{unno1989}. 
To avoid the Cowling approximation, $\mathcal J$ is used to modify the two characteristic frequencies \citep{takata2006}, considering only dipolar modes ($\ell = 1$), as
\begin{equation}
\tilde N = \frac{N} {\mathcal J} \,\,\, \text{and} \,\,\,  \tilde S_1 = {\mathcal J} S_1.
 \label{eq:modified}
\end{equation}
Rearranging equations~\eqref{eq:vr} to~\eqref{eq:modified}, $\tilde N$ and $\tilde S_1$ can be rewritten in terms of P and Q as 
\begin{equation}
\tilde N^2 = \frac{g}{r}  \frac{\mathcal A}{\mathcal J} = \frac{\omega^2 \left( \mathcal J - Q \right) }{\mathcal J}  
\label{eq:modified_N}
\end{equation}
and
\begin{equation}
\tilde S_1^2 = \frac{g}{r} \frac{2 \mathcal J}{\mathcal V} = \frac{2\omega^2\mathcal J}{2\mathcal J-P}.
 \label{eq:modified_S}
\end{equation}

\begin{figure}
\resizebox{\columnwidth}{!}{\includegraphics[angle =0]{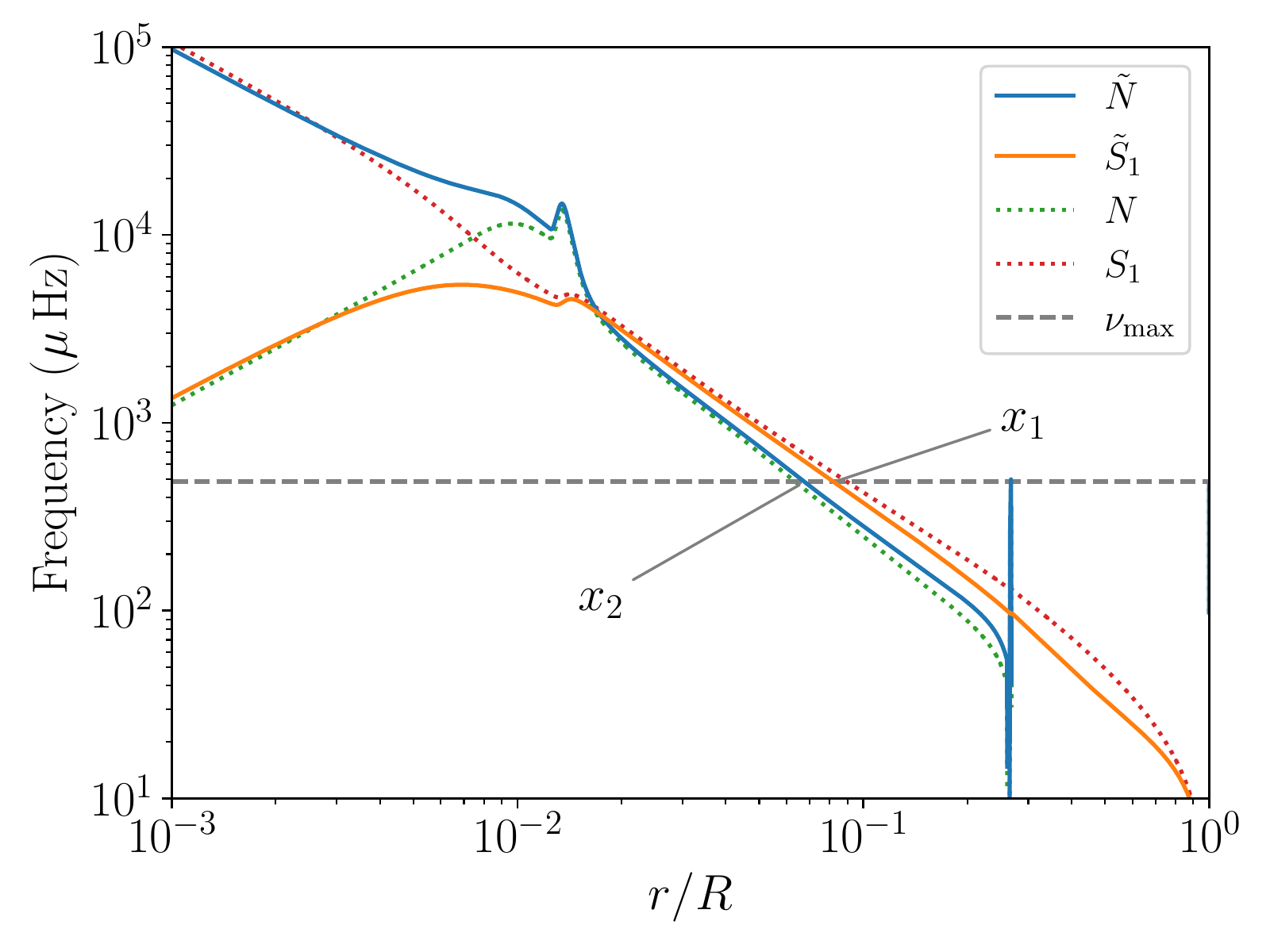}}
\caption{Comparison between the original characteristic frequencies ($N$ and $S_1$, dotted curves) and their modified versions ($\tilde N$ and $\tilde S_1$, solid curves) for a $1.0~\msun$, solar composition model. A smaller EVZ is produced by the modified characteristic frequencies for detectable oscillation modes with frequencies around the $\numax$ ($486.4\,\muHz$). The locations of the turning points $x_1$ and $x_2$ correspond to $\numax  = \tilde S_1$ and $\numax = \tilde N$, respectively.}
\label{fg:com_propagation}
\end{figure}

 \begin{figure}
\resizebox{\columnwidth}{!}{\includegraphics[angle =0]{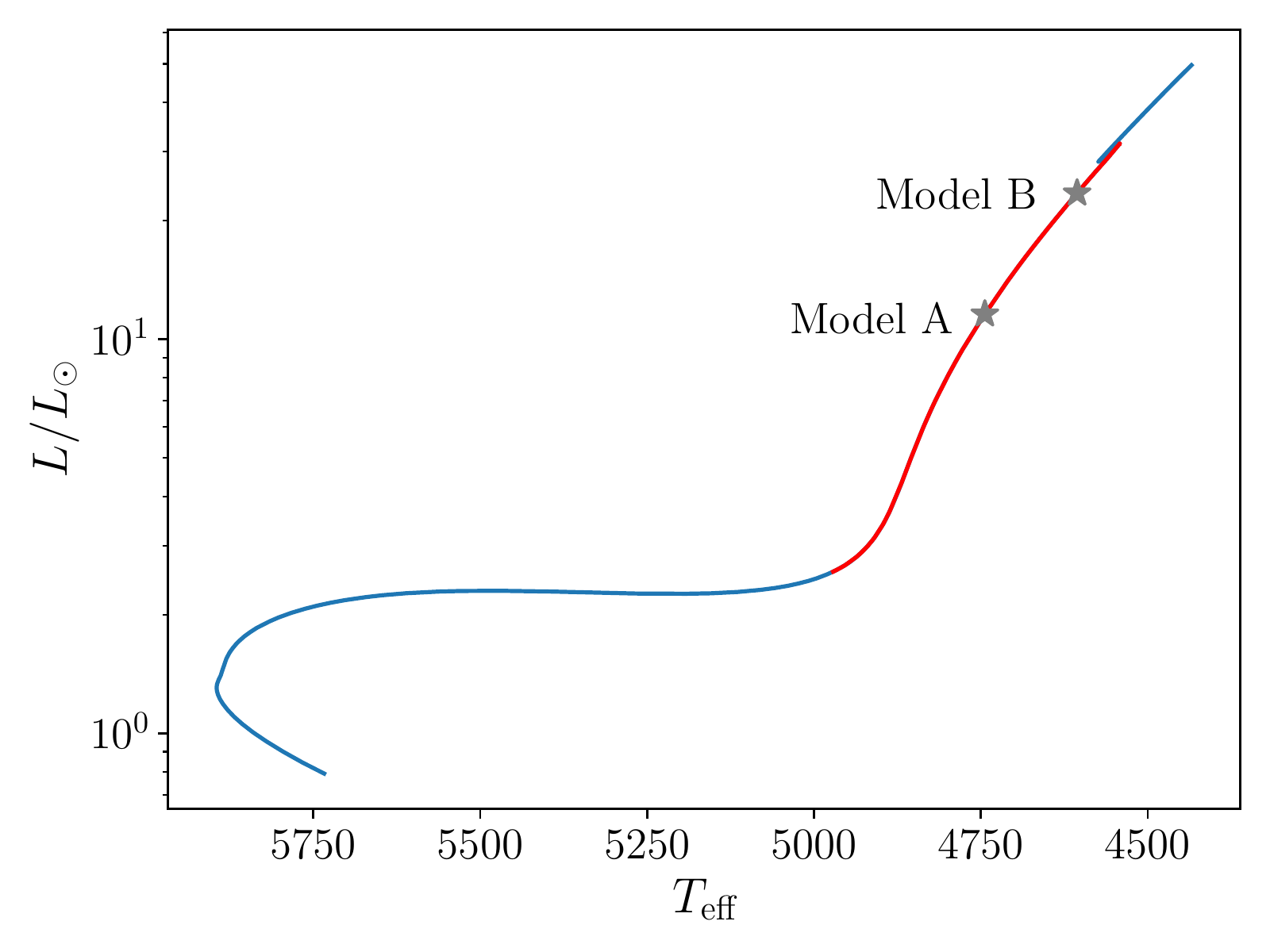}}
\caption{Evolution track of the 1.0 $\msun$, solar composition models. The red section are models with a buoyancy spike (glitch) present in the outer half the g-mode cavity (either at the outer edge or inside the g-mode cavity).  Two models highlighted by the star symbols are selected for further analyses (see main text). }
\label{fg:track}
\end{figure}

\begin{figure*}
\resizebox{\textwidth}{!}{\includegraphics[angle =0]{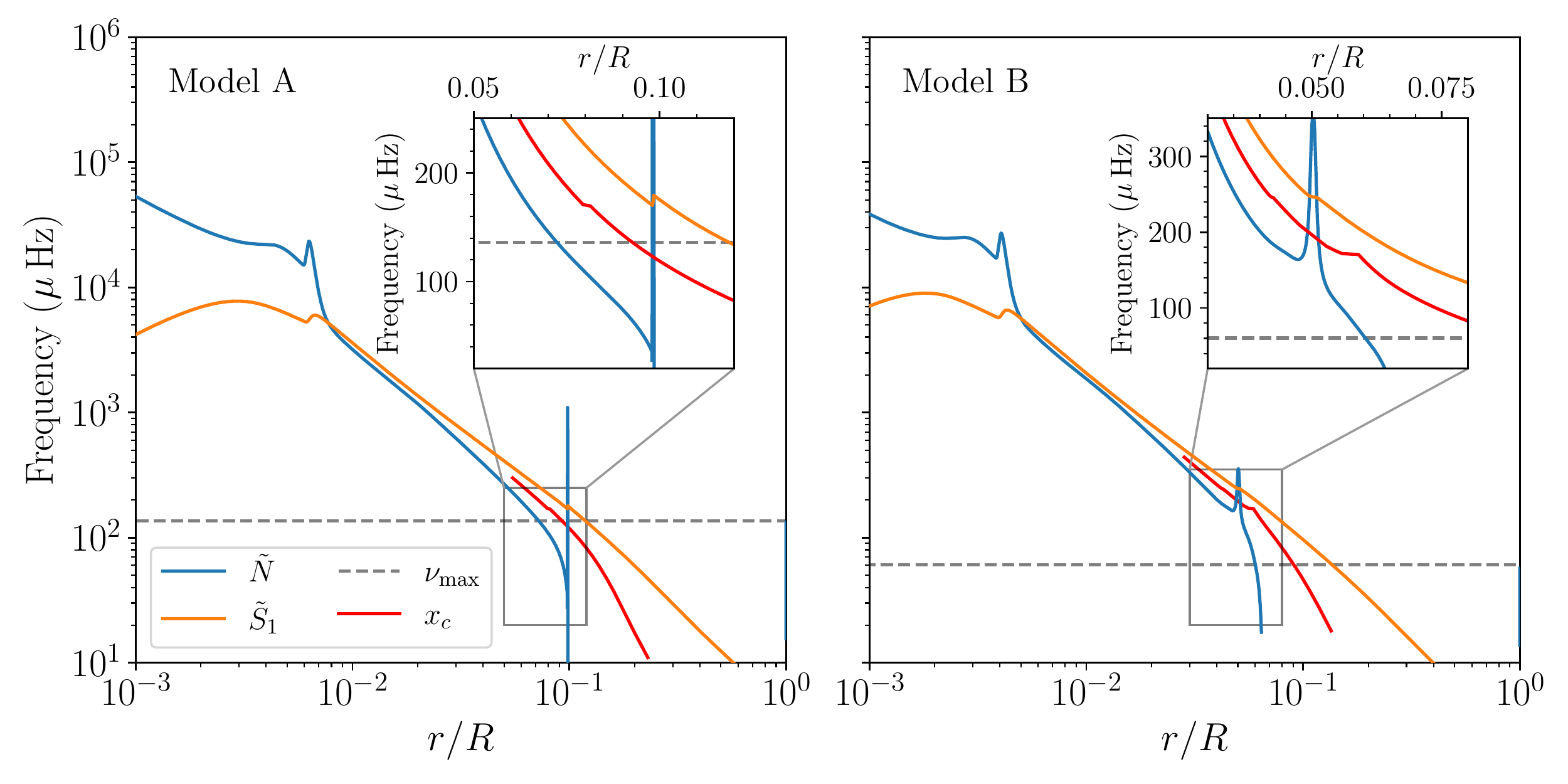}}
\caption{Propagation diagram illustrating $\tilde N$ and $\tilde S_1$ for Model A and Model B. The central positions of the EVZ, $x_c$, are computed as a continuous function of frequency. The insets zoom into the region of the EVZ in the vicinity of the buoyancy spike (glitch), where small displacements in $\tilde S_1$ and $x_c$ are seen for both models, along with the rapid buoyancy variations. }
\label{fg:propagation}
\end{figure*}

\begin{figure*}
\resizebox{\textwidth}{!}{\includegraphics[angle =0]{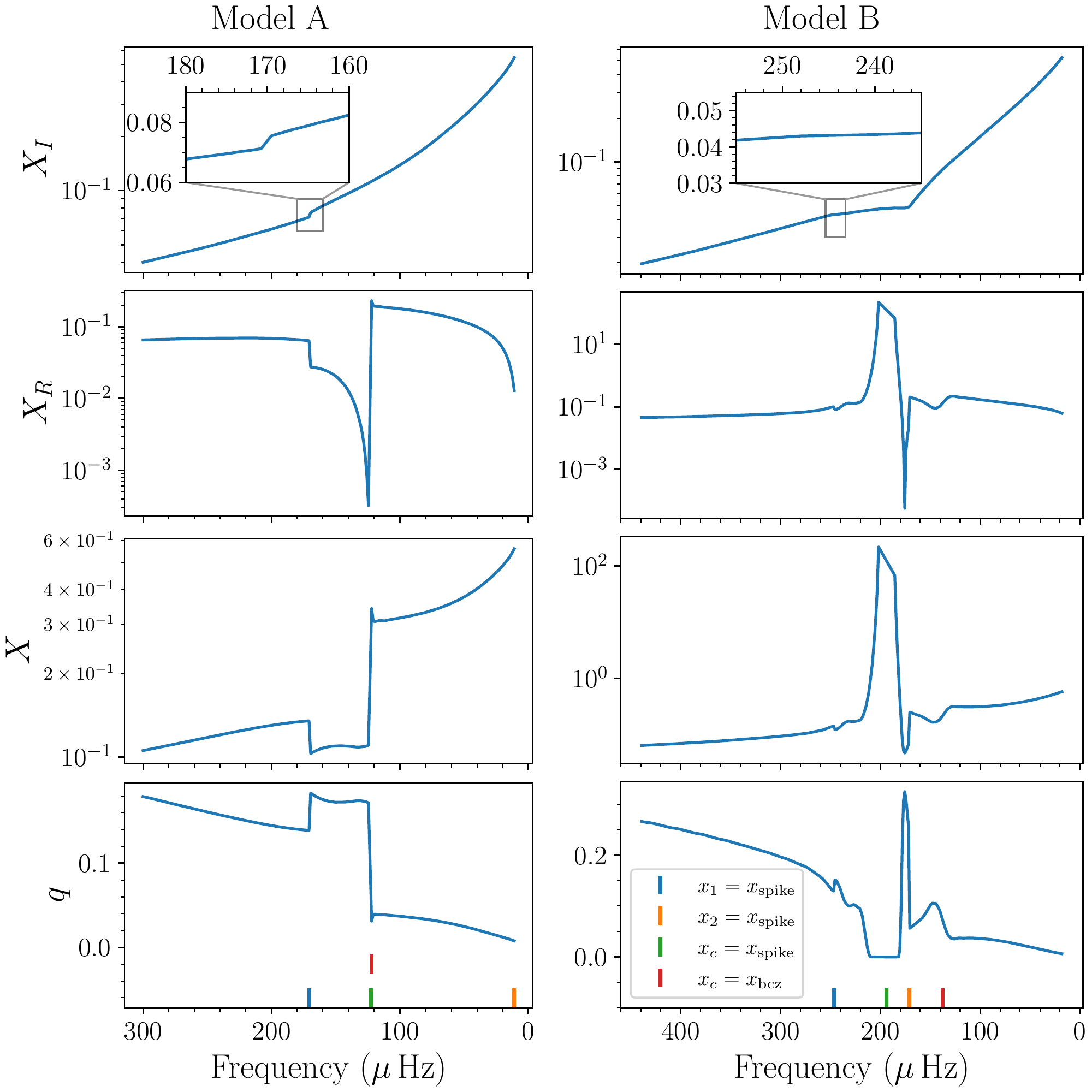}}
\caption{Variations of $X_I$, $X_R$, $X$ and the resulting asymptotic $q$ (with equation~\eqref{eq:qasy}) induced by the structural variations, shown as a continuous function of frequency, for both Model A and Model B. The coloured vertical bars in the bottom panels indicate the four critical positions where the turning points and $x_c$ coincide with the spike (or centre point of the glitch) and the BCZ in different cases (see the legend), as frequency decreases. The insects in the top panels show the subtle change in $X_I$ gradient when $x_1$ crosses the $x_{\rm spike}$. However, in the case of Model B, this change in $X_I$ is insignificant due to the broadening of the structural variations.}
\label{fg:var_q}
\end{figure*}

For the reason that $\mathcal J$ is below unity, $\tilde N$ and $\tilde S_1$ are expected to produce a slightly narrower EVZ than the normal counterparts do (see Figure~\ref{fg:com_propagation}). This difference diminishes for evolved red giant stars as $\mathcal J \approx 1$ in the EVZ. In the asymptotic limit of T16, the outer acoustic waves propagate in a cavity where $\omega > (\tilde S_1$ and $\tilde N$), while internal gravity waves are maintained by gravity in a central radiative cavity where $\omega < (\tilde N$ and $\tilde S_1$). Oscillation mode behaves exponentially in the EVZ, thus, is evanescent. The size of the EVZ is characterised by the outer and inner turning points that are the interfaces between the propagative cavities and the the EVZ, denoted as $x_1$ (outer) and $x_2$ (inner) in the unit of fractional radius $x = r/R$ with $R$ being the surface radius. An illustration of the locations of the turning points is given in Figrue~\ref{fg:com_propagation}.

All quantities in equations~\eqref{eq:vr} to~\eqref{eq:modified} are functions of $r$. T16 introduced an alternative independent variable $s$ that is connected with $x$ (and $r$) by
\begin{equation}
s = \ln x - \frac{1}{2} (\ln x_1 + \ln x_2) = \ln \frac{x}{\sqrt{x_1 x_2}}.
\end{equation}
The turning points $x = x_1$ and $x = x_2$ are then transformed to $s = s_0$ and $s = -s_0$, respectively, in which
\begin{equation}
s_0 = \frac{1}{2} (\ln x_1 - \ln x_2) = \ln \sqrt{\frac{x_1}{x_2}}
\end{equation}
Note that $s_0$ can be negative when $x_1 < x_2$, as seen in very young giant star. The centre of the EVZ ($x_c$) is defined as the point where $s = 0$, corresponding to $x_c = \sqrt{x_1 x_2}$. Since $x_1$ and $x_2$ are frequency dependent, the $s=0$ point and $x_c$ also change with frequency.

\begin{figure*}
\resizebox{\textwidth}{!}{\includegraphics[angle =0]{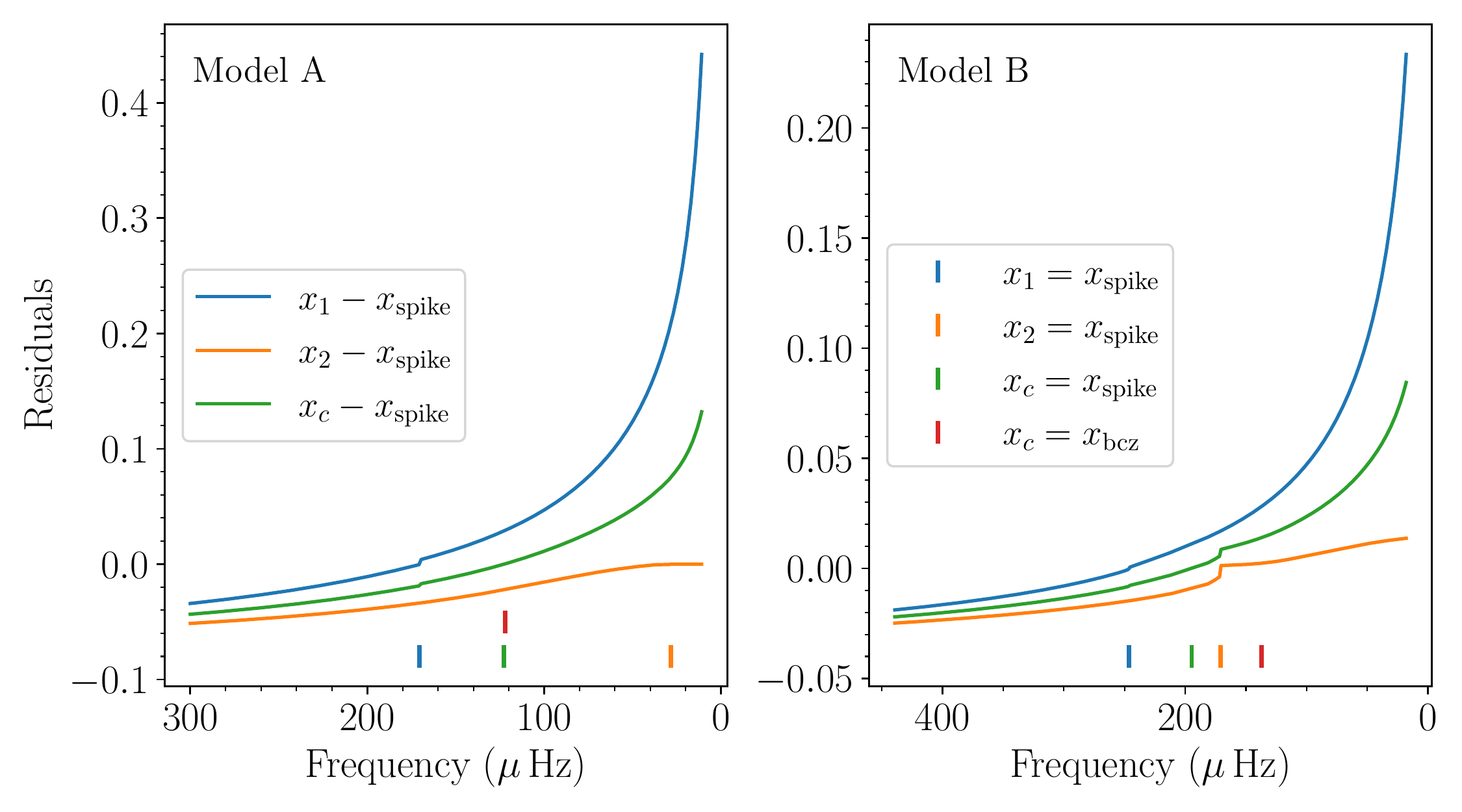}}
\caption{Locations of the turning points $x_1$, $x_2$ and the EVZ centre $x_c$, relative to the location of the spike (glitch) $x_{\rm spike}$, against the continuous frequency. The corresponding frequencies of the four critical positions are the zeros of each residual curve, indicated by the coloured vertical bars. In Model A, only subtle jumps in $x_1$ and $x_2$ are seen at $x_1 = x_{\rm spike}$. In Model B, $x_2$ and $x_c$ also jump to higher values at $x_2 = x_{\rm spike}$ due to the larger width of the buoyancy glitch compared to Model A. }
\label{fg:cric_fre}
\end{figure*}

With the independent variable $s$, $P$ and $Q$ are now transformed to 
\begin{equation} 
\frac{P}{F^2} = - S(s) (s-s_0)
\label{eq:ps}
\end{equation}
and
\begin{equation} 
Q{F^2} = T(s) (s+s_0),
\label{eq:qs}
\end{equation}
in which 
\begin{equation}
S > 0 ~~ \text{and} ~~ T > 0,
\label{eq:st_condition}
\end{equation}
and
\begin{equation}
F(x) = \exp{ \left[ \frac{1}{2} \int^x \frac{ \mathcal{V}(x_{\rm a}) - \mathcal{A}(x_{\rm a}) - \mathcal{J}(x_{\rm a}) }{x_{\rm a}} \mathrm d  x_{\mathrm a} \right]}.
\end{equation}
The turning points $s_0$ and $-s_0$ are indeed the zeros of $P$ and $Q$, respectively. 

According to T16, in the limiting case of a thin EVZ, the coupling factor is asymptoticly given by
\begin{equation} 
q = \frac{1-\sqrt{1-\exp{(-2X)}}}{1+\sqrt{1-\exp{(-2X)}}},
\label{eq:qasy}
\end{equation}
where $X$ is a positive parameter consisting of two terms as
\begin{equation}
X = X_I + X_R.
\end{equation}
$X_I$ is an integral term similar to the one on the right-hand side of equation~\eqref{eq:qweak}, given by
\begin{equation}
X_I = \int_{-|s_0|}^{|s_0|} \kappa(s) \sqrt{s_0^2 - s^2} \mathrm d s, 
\label{eq:xi}
\end{equation}
where $\kappa$ is now the wavenumber determined by the modified characteristic frequencies, and defined in terms of $P$ and $Q$ as a function of $s$ by
\begin{equation}
\kappa(s) = \sqrt{S T}  = \sqrt{\frac{-P Q} {s^2 - s_0^2}}.
\end{equation}
And $X_R$ is a gradient term that is sensitive to the variations of $\tilde N$ and $\tilde S_1$ in the vicinity of $x_{\rm c}$ where $s=0$. The definition of $X_R$ is
\begin{equation}
X_R = \frac{1}{2 \uppi \kappa (s = 0)} \left(  \frac{\mathrm d \ln \mathfrak c}{\mathrm d s} \right)^2_{s=0},
\label{eq:xr}
\end{equation}
in which
\begin{equation}
\mathfrak c = \left(\frac{S}{T}\right)^{1/4} = \frac{1}{F} \left( \frac{-P}{s-s_0} \frac{s+s_0}{Q} \right)^{1/4}.
\label{eq:c}
\end{equation}
The value of $\kappa$ and $\mathfrak c$ are obtained from equations~\eqref{eq:ps} and~\eqref{eq:qs}. Therefore, the stellar structure near the centre of the EVZ is critical to $q$. $Q$ and $P$ can be converted to $\tilde N$ and $\tilde S_1$ with equations~\eqref{eq:modified_N} and ~\eqref{eq:modified_S}, thus, $q$ is closely related to the properties of $\tilde S_1$ and $\tilde N$ over the region encompassing the EVZ in the interval $-|s_0|$ to $|s_0|$.

Since the structural variations start to emerge in the mid-layers in early RGB stars where a thin or intermediate--size EVZ exists, the following analysis is under the asymptotic framework of T16\footnote{Note that for an intermediate size (not very thick or very thin) EVZ, the asymptotic formalism in T16 may not converge with observations \citep{pincon2020}, due to its limiting case of a very thin EVZ. However, the asymptotic analysis of T16 is still helpful to investigate the relation between mid-layer structure and mixed-mode property.}. Furthermore, hereafter the modified version of the characteristic frequency are used but, for simplicity, they are called by their original names, i.e. buoyancy frequency and Lamb frequency. 

\begin{figure*}
\resizebox{\textwidth}{!}{\includegraphics[angle =0]{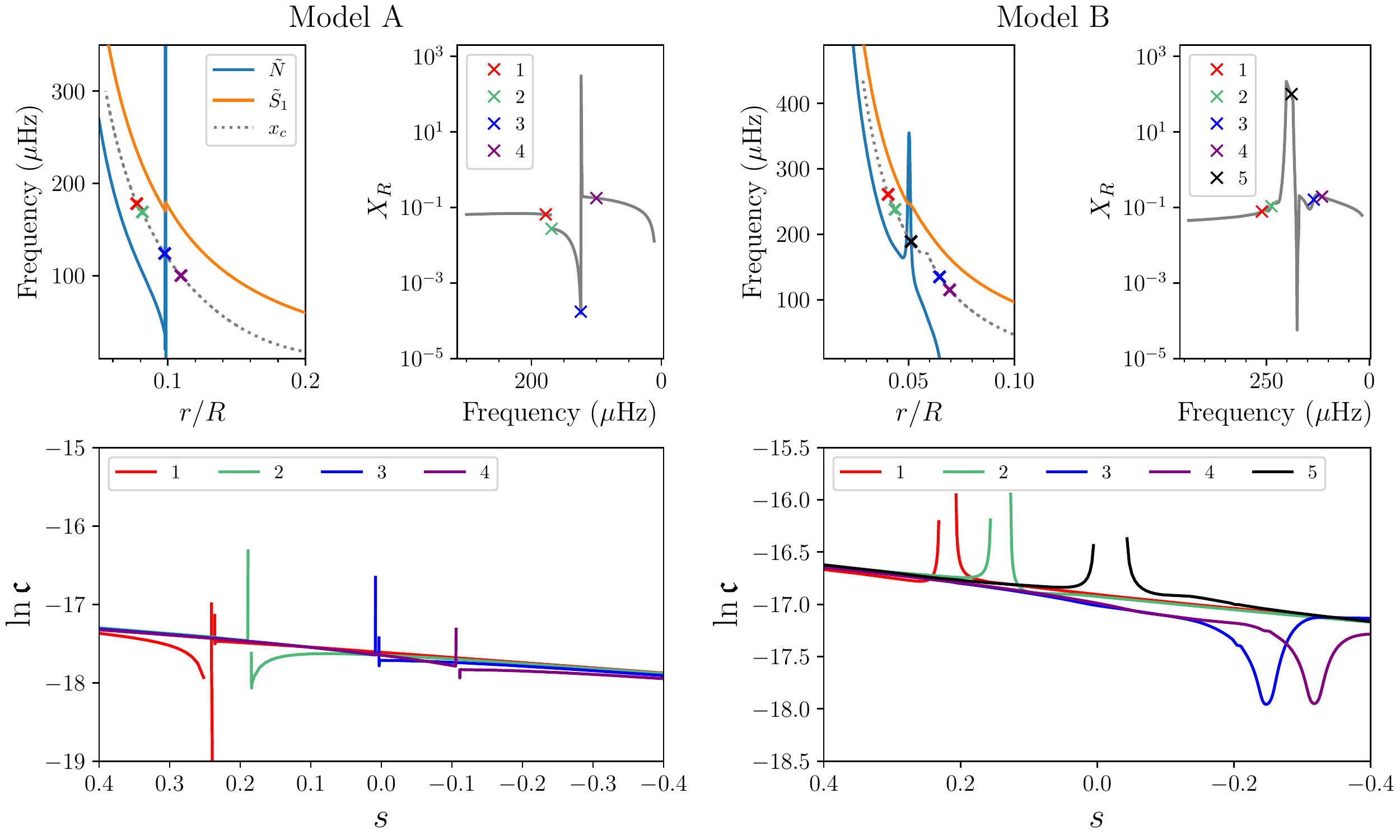}}
\caption{$\ln \mathfrak{c}$ as a function of $s$ (lower row) for several selected modes in a region encompassing the EVZ, for Model A and Model B. The locations of the selected modes in the propagation diagram and their $X_R$ values are indicated by the cross symbol in the upper row panels. The sharp variations in $\ln \mathfrak{c}$ are mainly generated by the Lamb frequency displacement and the buoyancy spike (glitch). The stellar depth increases as $s$ decreases in the abscissa.}
\label{fg:lnc_ori}
\end{figure*}

\begin{figure*}
\resizebox{\textwidth}{!}{\includegraphics[angle =0]{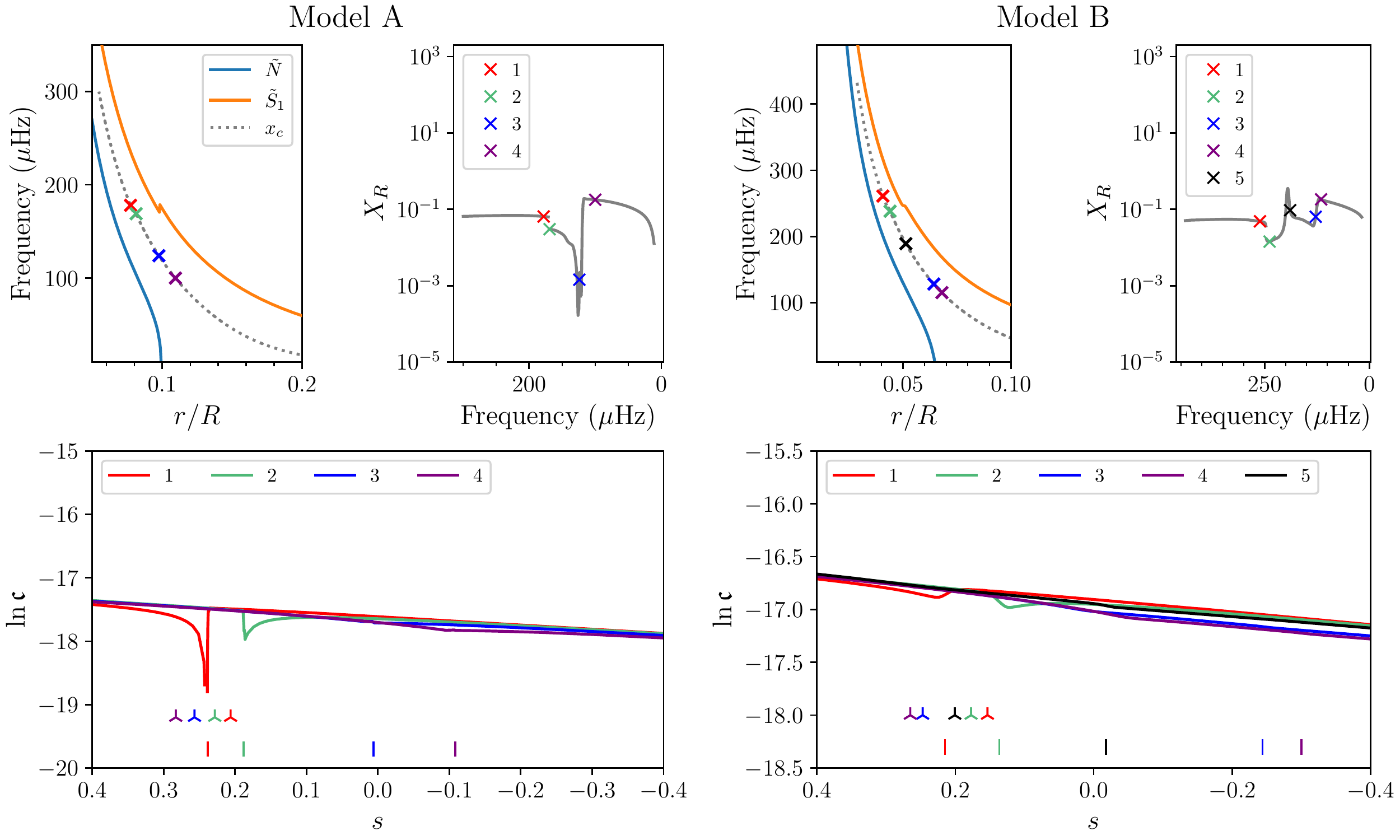}}
\caption{$\ln \mathfrak{c}$ as a function of $s$ for modes with the same frequencies as in Figure~\ref{fg:lnc_ori}. The buoyancy spike (glitch) is artificially removed in Model A (B) so that the significant sharp variations in $\ln \mathfrak{c}$ are only caused by the $\tilde S_1$  displacement whose positions are indicated by the coloured vertical bars. And the flipped 'Y' symbols are the outer turning points $s_0$. The colours of symbols are consistent with the colours of the modes. Note the position of the $\tilde S_1$ displacement is also the position of the removed buoyancy spike (glitch), as they are both caused by the structural variations.}
\label{fg:lnc_s1}
\end{figure*}

\subsection{Variations of Asymptotic Coupling Factor}
\label{sc:varq}

Stellar models computed with \texttt{ASTEC} \citep{jcd08a} and corresponding oscillations with \texttt{ADIPLS} \citep{jcd08b} are used in this analysis. The models have an initial mass of $1.0~\msun$, and a heavy-element abundance of 0.0173 (calibrated to solar model). A mixing-length parameter of 1.96 (calibrated to solar model) is employed for the treatment of convection under the assumption of the mixing-length theory \citep{bohm1958}. The models are kept as simple as possible, meaning no overshooting, diffusion or rotation are considered in the generation of the models. An illustration of the evolutionary track is shown in Figure~\ref{fg:track}.

Two models along the evolutionary track are selected to represent the case  of a buoyancy spike emerging at the BCZ (Model A, $\numax = 135.8 \, \muHz$), and the case of a buoyancy glitch penetrating inside the g-mode cavity (Model B, $\numax = 60.4 \, \muHz$). Recalling that the buoyancy spike (glitch) is caused by the structural variations, the location of the buoyancy spike (glitch) is indeed the location of the structural variations. The propagation diagrams of these two models are shown in Figure~\ref{fg:propagation}. Moreover, in Model A, the sharp buoyancy spike is located within the EVZ of detectable modes with frequencies near the $\numax$, which inevitably affect the inferred asymptotic $q$ by equation~\eqref{eq:qasy}. On the other hand, in Model B, given the fact that the buoyancy glitch is not inside the EVZ of detectable modes, variation in asymptotic $q$ is thus not expected to be observed for these modes\footnote{This is not necessarily true for observed coupling factor. The glitch has an impact on the oscillation modes as long as it is inside the g-mode cavity, starting from low-order mode with frequencies much smaller than $\numax$. This influence can also be felt by modes in detectable frequency range, though with smaller scale of perturbations \citep{jiang2022}.}, instead it can be seen in lower-frequency mixed modes.

Along with the rapid variations in $\tilde N$, the structural variations also induce a subtle displacement in $\tilde S_1$ that is exhibited in the insets of Figure~\ref{fg:propagation}. Consequently, $x_1$ would experience a sudden jump when the outer turning point crosses the $\tilde S_1$ displacement as the mode frequency decreases, which further leads to the little jump in $x_c$ (see the insects of Figure~\ref{fg:propagation} for details). These variations in the characteristic frequencies break the assumption of smooth change in the asymptotic analysis. As a result, the inferred asymptotic $q$ will deviate from the case that is free of structural variations. Figure~\ref{fg:var_q} shows how $X_I$, $X_R$, $X$ and the resulting $q$ change with the frequency for both Model A and Model B. In both cases, significant variations in the four quantities are seen at four critical positions where $x_1$, $x_2$ and $x_c$ respectively meet the location of the spike (glitch), $x_{\rm spike}$\footnote{For simplicity, the location of the buoyancy glitch is also represented by $x_{\rm spike}$.}, and that of the BCZ, $x_{\rm bcz}$, as frequency decreases. The locations of the critical positions and their corresponding frequencies are shown Figure~\ref{fg:cric_fre}. In particular, the models in Figure~\ref{fg:var_q} highlight the following points:
\begin{itemize}
\item[--] In both models, $X_I$ shows a tiny change in its variation with respect to frequency at the position of $x_1=x_{\rm spike}$. Based on equation~\eqref{eq:xi}, $X_I$ is directly related to the values of $\kappa$ enclosed in the EVZ as well as the size of the EVZ determined by the locations of the turning points $s_0$ and $-s_0$. The value of $s_0$ would have a sudden increase when the outer turning point passes the displacement in $\tilde S_1$ resulted from the structural variations. This in combination with the fact that the value of $\kappa$ is also influenced by the displacement in $\tilde S_1$, leads to the tiny change in $X_I$. However, in Model B, this change in $X_I$ is insignificant, which is due to the fact that the width of the structural variation of Model B is relatively large so the $\tilde S_1$ displacement, hence the inferred $X_I$, is not sharp but rather smooth, compared with Model A.
\item[--] $X_R$ is sensitive to $\kappa$ as well as the gradient of $\ln \mathfrak{c}$, hence the gradient of $\tilde N$ and $\tilde S_1$, at the center of the EVZ. In Model A, $X_R$ has two significant jumps at $\sim 170$ and $125\,\muHz$, corresponding to the critical positions of $x_1 = x_{\rm spike}$ and $x_c = x_{\rm spike}$, respectively. The first jump is also caused by the displacement in $\tilde S_1$ when $x_1$ passes $x_{\rm spike}$ overlapping with $x_{\rm bcz}$ (details given in Section~\ref{sc:lamb}). The second jump occurs as the buoyancy spike approaches $x_c$ where $X_R$ is calculated. Thus, the second jump has much larger scale of variations, compared with the first jump for which the change in $\kappa$ only induces small variations at $x_c$.
\item[--] In Model B, $X_R$ also experiences two jumps as Model A, one small jump located at the critical position of $x_1 = x_{\rm spike}$ and one with larger scale at $x_c = x_{\rm spike}$ (variations also exist in their vicinities due to the rather large width of the glitch), respectively. Furthermore, additional variations in $X_R$ are present around the position of $x_2 = x_{\rm spike}$ where $x_2$ and $x_c$ undergo a jump (see Figure~\ref{fg:cric_fre} and the caption) due to the glitch; and at the position of  $x_c = x_{\rm bcz}$ where $\tilde N$ and $\tilde S_1$ start to vary differently with radius near the BCZ, in contrary to the relatively parallel variations in deeper layers. While $x_2 > x_{\rm spike}$, meaning the glitch has moved out of the EVZ as frequency decreases, the value of $X_R$ resume the normal level that is free of glitch impact. However, as for Model A, there is no significant jump in $X_R$, and other three quantities at $x_2 = x_{\rm spike}$, because the buoyancy spike is so sharp that $x_2$ is not affected but varies smoothly with frequency at the position of the spike (see Model A case of Figure~\ref{fg:cric_fre}).
\item[--] Since $X$ is the sum of $X_I$ and $X_R$, all variations in $X_I$ and $X_R$ discussed above will propagate to $X$. In general, $X$ is dominated by $X_R$ in high-frequency range corresponding to the case of low-luminosity RGB stars in which $\numax$ are high and the EVZ is inside the radiative region, and is dominated by $X_I$ in low-frequency range corresponding to evolved RGB stars in which $\numax$ are low and the EVZ is in the convective region. In intermediate--frequency range where significant changes due to the structural variations are seen near the BCZ, $X_I$ and $X_R$ are comparable. 
\item[--] As a result, the value of $q$ changes according to the variations in $X$ at the critical positions. In summary, the structural variations induce sharp changes in $\tilde N$ and a small displacement in $\tilde S_1$ that further cause the changes in asymptotic $q$ calculated under the assumption that the model is free of structural variations. Compared to the displacement in $\tilde S_1$, the buoyancy spike (glitch) gives rise to larger scale variations of $q$. Moreover, the different variations of the buoyancy and Lamb frequencies with radius near the BCZ bring about additional contributions to the jump in $q$ as the EVZ is centred at the BCZ, and are the reason of the quick drop in $q$ when the EVZ is entirely above the BCZ, which can be observed in evolved RGB stars \citep{hekker2017}. However, when $x_{\rm spike}$ overlaps with $x_{\rm bcz}$, as in Model A and low-luminosity RGB stars, the contributions of the buoyancy spike and the BCZ to the variations of $q$ cannot be disentangled. On contrary, when the spike becomes a glitch and moves away from the BCZ to be located inside the g-mode cavity, as in Model B and evolved RGB stars, the contributions of BCZ is unfolded.
\end{itemize}
 
\section{Contributions to the variations of asymptotic Coupling factor}
\label{sc:contribution}

The asymptotic $q$ from equation~\eqref{eq:qasy} is dependent on the value of $X$ that is the sum of $X_I$ and $X_R$. Since $X_I$ only has a small displacement at the position of the spike (glitch) and its contribution to the variations of the inferred $q$ is straightforward to understand, in this section we specifically discuss the different contributing factors to the jumps in $X_R$ and $q$. Bearing this in mind, several modes with different frequencies are selected to illustrate how values of $X_R$, and hence $q$, are impacted at the critical positions (Figure~\ref{fg:lnc_ori}). For Model A, as the mode frequency decreases there are two major jumps in $X_R$ at the positions where the mode with a certain frequency satisfies $x_1 = s_{\rm spike}$ for one jump and $x_c = x_{\rm bcz}$ for the other. Thus, four modes are selected to represent the cases that the mode has a frequency just higher or lower than the frequencies corresponding to the two critical positions. For instance, as shown in Figure~\ref{fg:lnc_ori}, Modes 1 and 2 selected for Model A are respectively on the left and right side of the first jump in $X_R$ at $\sim 170 \,\muHz$, and the rest two modes (Modes 3 and 4) are on the sides of the second jump at $\sim 125\,\muHz$. 
For Model B, in additional to the four modes (Modes 1-4) illustrating the cases around the critical positions of $x_1 = s_{\rm spike}$ and $x_c = x_{\rm bcz}$ as in Model A, one mode (Mode 5) is also selected to depict the case of $x_c = x_{\rm spike}$. In this way, we are able to examine explicitly how the quantity $\ln \mathfrak{c}$ differs between the modes, and particularly the value of $\ln \mathfrak{c}$ at the centre of the EVZ, which is crucial to the determination of $X_R$. 

The panels in the first row of Figure~\ref{fg:lnc_ori} give the locations of the selected modes for both Model A and Model B, while the panels in the second row show $\ln \mathfrak{c}$ as a function of $s$ in a region encompassing the EVZ with the centre of the EVZ being $s=0$. The values of $\ln \mathfrak{c}$ progressively decrease in depth as a straight line in regions that are free of the impact of the structural variations. Significant sharp variations are present for the otherwise linear decreasing $\ln \mathfrak{c}$ with depth, for all the modes. The width of the sharp variations is in scale with the width of the buoyancy spike (glitch). The gap seen within the central part of the sharp variations (clear in Model B) is due to the fact that the spike (glitch) leads to negative $Q$ and $T$ in this area, which is against equation~\eqref{eq:st_condition}. Therefore, $\mathfrak{c}$ is not real in this area according to equation~\eqref{eq:c}.

As discussed in Section~\ref{sc:varq}, the sharp variations in $X_R$ and $\ln \mathfrak{c}$ are expected to be mainly caused by the buoyancy spike (glitch) in Model A (B), and partially by the displacement in the Lamb frequency as well as the gradient change of the characteristic frequencies near the BCZ. In order to distinguish the contributions from the three factors, we modify the profile of $\tilde N$ and $\tilde S_1$ by using a Python utility\footnote{\url{https://github.com/stefano-rgc/glitch_add_and_remove}} that allows for artificially removing the structural variations induced changes in the characteristic frequencies from the models. In the following subsections, the three factors are investigated separately with the modified models.

\begin{figure}
\resizebox{\columnwidth}{!}{\includegraphics[angle =0]{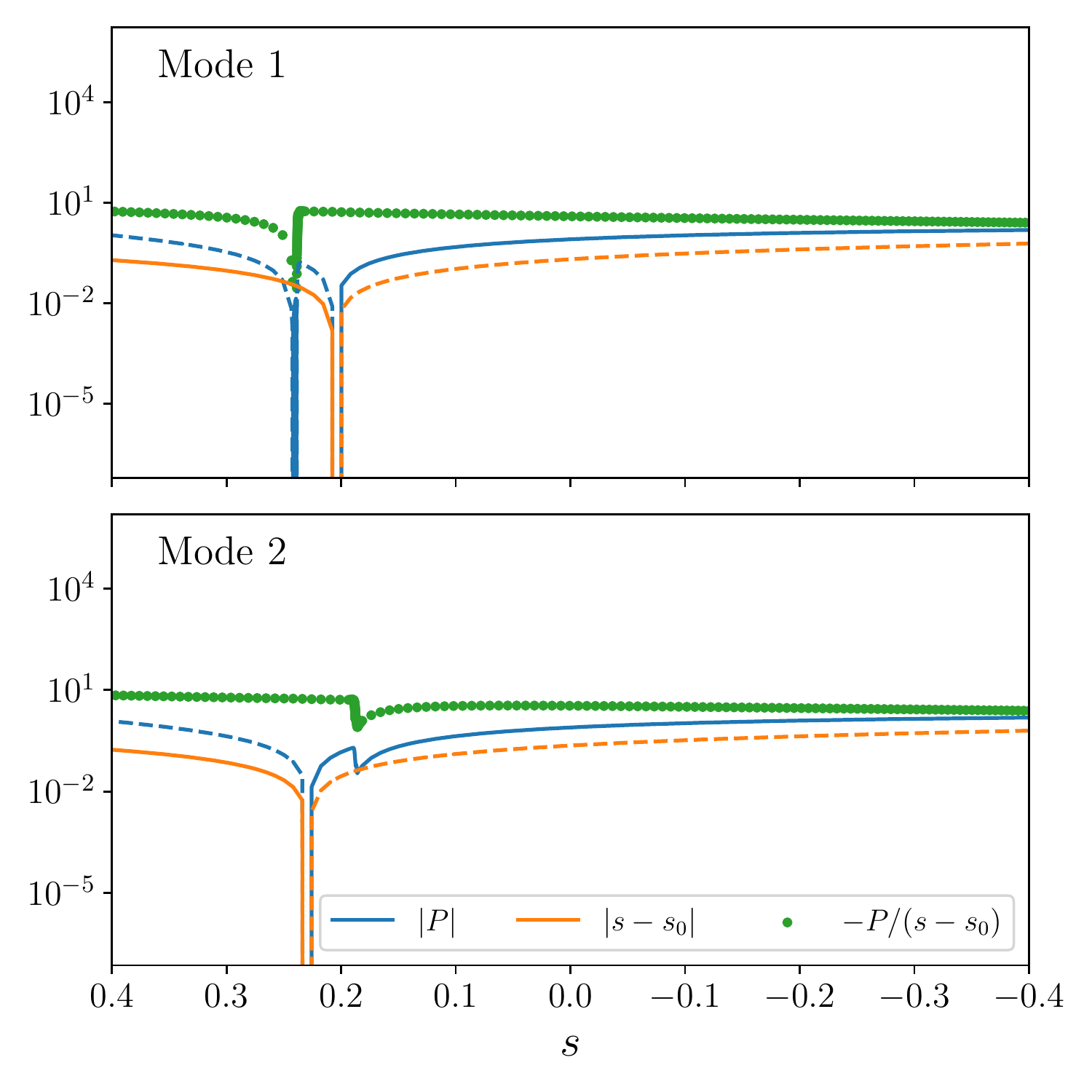}}
\caption{Values of $|P|$ (blue), $|s-s_0|$ (orange) and the resulting $-P / (s-s_0)$ (green) for Mode 1 (upper) and Mode 2 (bottom) of Model A with the buoyancy spike (glitch) removed from the model. 
The singularity seen in $|s-s_0|$, overlapping the drop in $|P|$, indicates the location of $s_0$, while significant decrease happens at $s_{\rm spike}$.
A real value of $\mathfrak{c}$ requires $-P / (s-s_0) > 0$ (c.f. equations~\eqref{eq:st_condition} and~\eqref{eq:c}), thus different signs between $P$ and $s-s_0$. The negative parts of $P$ and $s-s_0$ are indicated by the dashed lines. }
\label{fg:comp_p}
\end{figure}

\begin{figure*}
\resizebox{\textwidth}{!}{\includegraphics[angle =0]{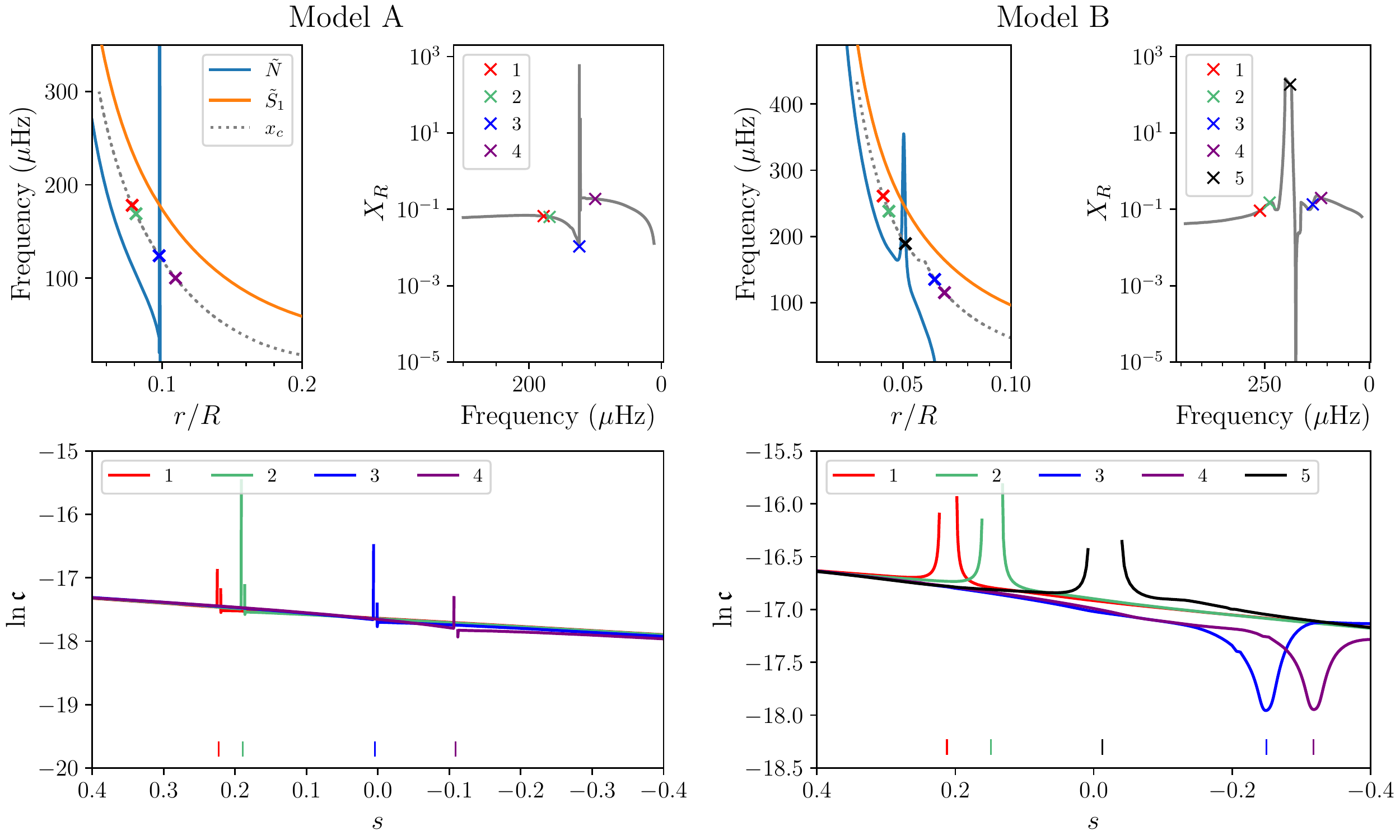}}
\caption{$\ln \mathfrak{c}$ as a function of $s$ for modes with the same frequencies as in Figure~\ref{fg:lnc_ori}. The displacement in the Lamb frequency is artificially removed in both Model A and Model B so that the significant sharp variations in $\ln \mathfrak{c}$ are primarily caused by the buoyancy spike (glitch) whose positions are indicated by the colored vertical bars. The colors of bars are consistent with the colors of the modes.}
\label{fg:lnc_ng}
\end{figure*}

\subsection{Contribution of Lamb Frequency Displacement}
\label{sc:lamb}

The contribution of the displacement in $\tilde S_1$ to $X_R$, and hence to the asymptotic $q$, is explicitly displayed in Figure~\ref{fg:lnc_s1}, when the buoyancy spike (glitch) is artificially removed from the stellar models. Compared to the original models in Figure~\ref{fg:lnc_ori}, most sharp variations in $\ln \mathfrak{c}$ are diminished and only Mode 1 and Mode 2 each has one significant sharp variation at the position of the spike (glitch) in $s$, $s_{\rm spike}$, defined as
\begin{equation}
s_{\rm spike} = \ln x_{\rm spike} - \frac{1}{2} (\ln x_1 + \ln x_2).
\end{equation}
As mentioned before, the difference between Mode 1 and Mode 2 is that the outer turning point $s_0 < s_{\rm spike}$ for Mode 1, but $s_0 > s_{\rm spike}$ for Mode 2, as indicated by the colored symbols in Figure~\ref{fg:lnc_s1}. This, together with the fact that $s_0$ (and $-s_0$) is a singularity of $\mathfrak{c}$, produces the different shape of the two sharp variations in $\ln \mathfrak{c}$. According to equation~\eqref{eq:c}, $\ln \mathfrak{c}$ is determined by $F$, $P$, $Q$ and $s_0$. $F$ is dominated by the properties near the stellar centre and thus is nearly constant in the region encompassing the EVZ. $Q$ is spikeless since the spike is removed from $\tilde N$, and thus varies smoothly with respect to $s$. Therefore, it is the property of $P$ and $s_0$ that governs the shape of the variations in $\ln \mathfrak{c}$. 
 A real value of $\mathfrak{c}$ requires $-P / (s-s_0) > 0$, thus different signs between $P$ and $s-s_0$.
The distribution of the detailed values of $|P|$, $|s-s_0|$ and the resulting $-P / (s-s_0)$ can help us understand the behaviour of the different shape of variations in $\ln \mathfrak{c}$ for Mode 1 and Mode 2, as shown in Figure~\ref{fg:comp_p} for Model A from which the buoyancy spike is removed. 
Since $s_0$ is a singularity of $P$, $P$ varies nearly in parallel with $(s-s_0)$ in depth, leading to the constant variation of  $-P / (s-s_0)$ and $\ln \mathfrak{c}$, except in the vicinity of $s_{\rm spike}$. 
The existence of the spike drastically changes the gradient of $P$ with respect to $s$, which produces not only the sharp variation in $\ln \mathfrak{c}$, but also shapes the variation according to the position of $s_0$ relative to $s_{\rm spike}$. For instance, in Mode 1 $s_0 < s_{\rm spike}$, so $-P / (s-s_0)$ decreases progressively in the region above the spike (recalling here the stellar depth increases as $s$ decreases) till $s_{\rm spike}$ where it resumes constant variation. 
On the other hand, in Mode 2 $s_0 > s_{\rm spike}$, the progressive change in $-P / (s-s_0)$ happens in layers deeper than the spike. 
This behaviour of $-P / (s-s_0)$ also propagates to $\ln \mathfrak{c}$ and $q$.

Another feature of the two sharp variations of Mode 1 and Mode 2 is that they appear below the straight lines of $\ln \mathfrak{c}$, meaning that the $\tilde S_1$ displacement reduces the value of $\ln \mathfrak{c}$ around $s_{\rm spike}$. The different shapes of the two sharp variations lead to further deviations at $s=0$ where $X_R$ is actually derived. At the $s=0$ point, the value of $\ln \mathfrak{c}$ of Mode 1 returns to the level of the straight lines that are not affected by the spike, but it is slightly reduced for Mode 2. This is the reason for the jump in $X_R$ located between Mode 1 and Mode 2 in both Model A and Model B. 
%However, the second jump between Mode 3 and Mode 4 is mainly caused by the gradient change of the characteristic frequencies at the BCZ, as discussed in Section~\ref{sc:bcz}. 
The value of $\ln \mathfrak{c}$ of Mode 3 and Mode 4 is also reduced by the displacement in $\tilde S_1$ at $s_{\rm spike}$, though in very small scale. Nevertheless, this can result in significant changes in $X_R$, hence $q$, if the spike is in the central of the EVZ, as in the case of Mode 3 in Model A as well as Mode 5 in Model B.

\begin{figure}
\resizebox{\columnwidth}{!}{\includegraphics[angle =0]{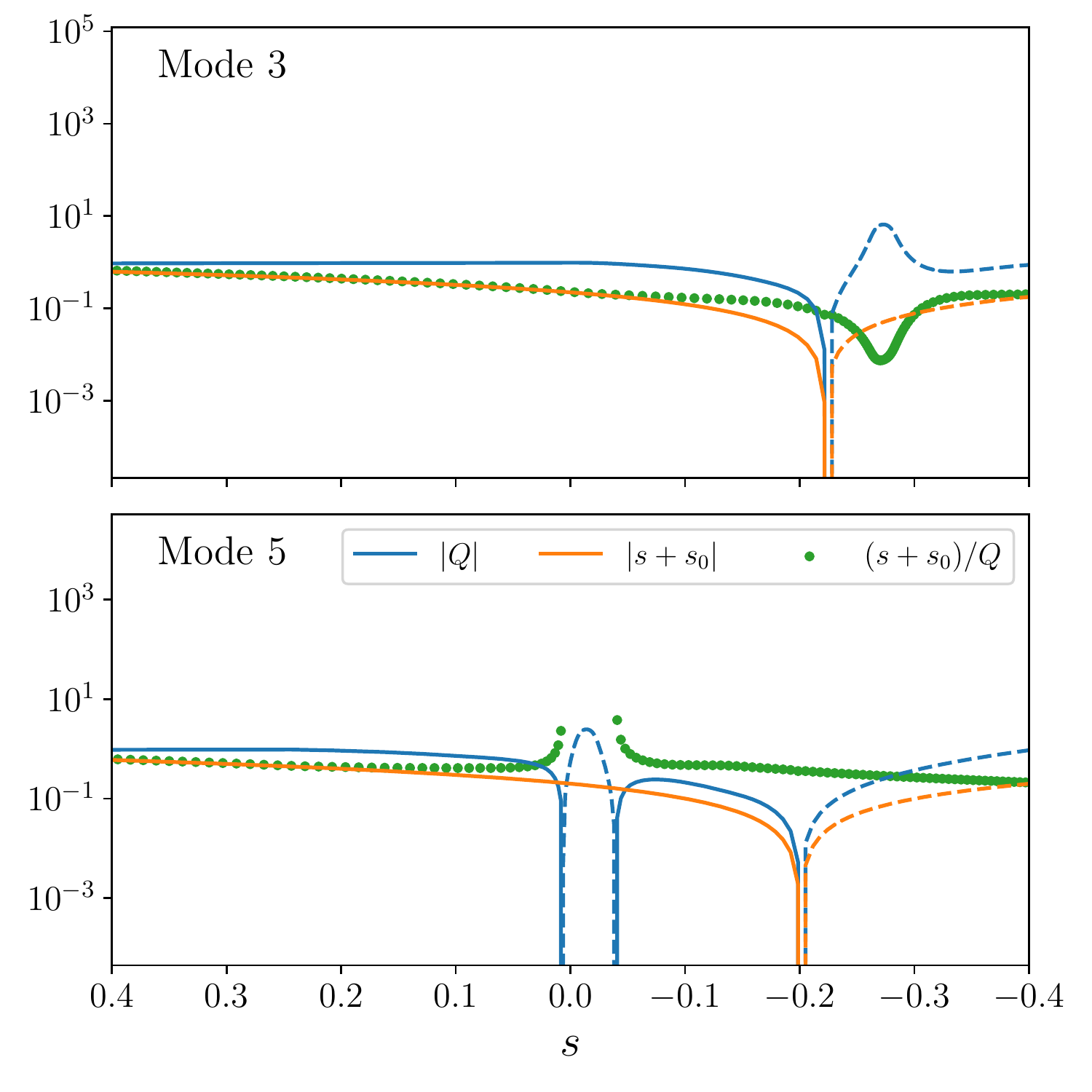}}
\caption{Values of $|Q|$ (blue), $|s+s_0|$ (orange) and the resulting $(s+s_0)/Q$ (green) for Mode 3 (upper) and Mode 5 (bottom) of Model B with the displacement in the Lamb frequency removed from the model. A real value of $\mathfrak{c}$ requires $(s+s_0) / Q > 0$ (c.f. equations~\eqref{eq:st_condition} and~\eqref{eq:c}), thus same signs between $Q$ and $s+s_0$. The gap around $s=0$ is where $(s+s_0) / Q < 0$, hence imaginary $\mathfrak{c}$. The negative parts of $Q$ and $s+s_0$ are indicated by the dashed lines. The glitch is outside the EVZ for Mode 3 but inside the EVZ for Mode 5.}
\label{fg:comp_q}
\end{figure}

\begin{figure*}
\resizebox{\textwidth}{!}{\includegraphics[angle =0]{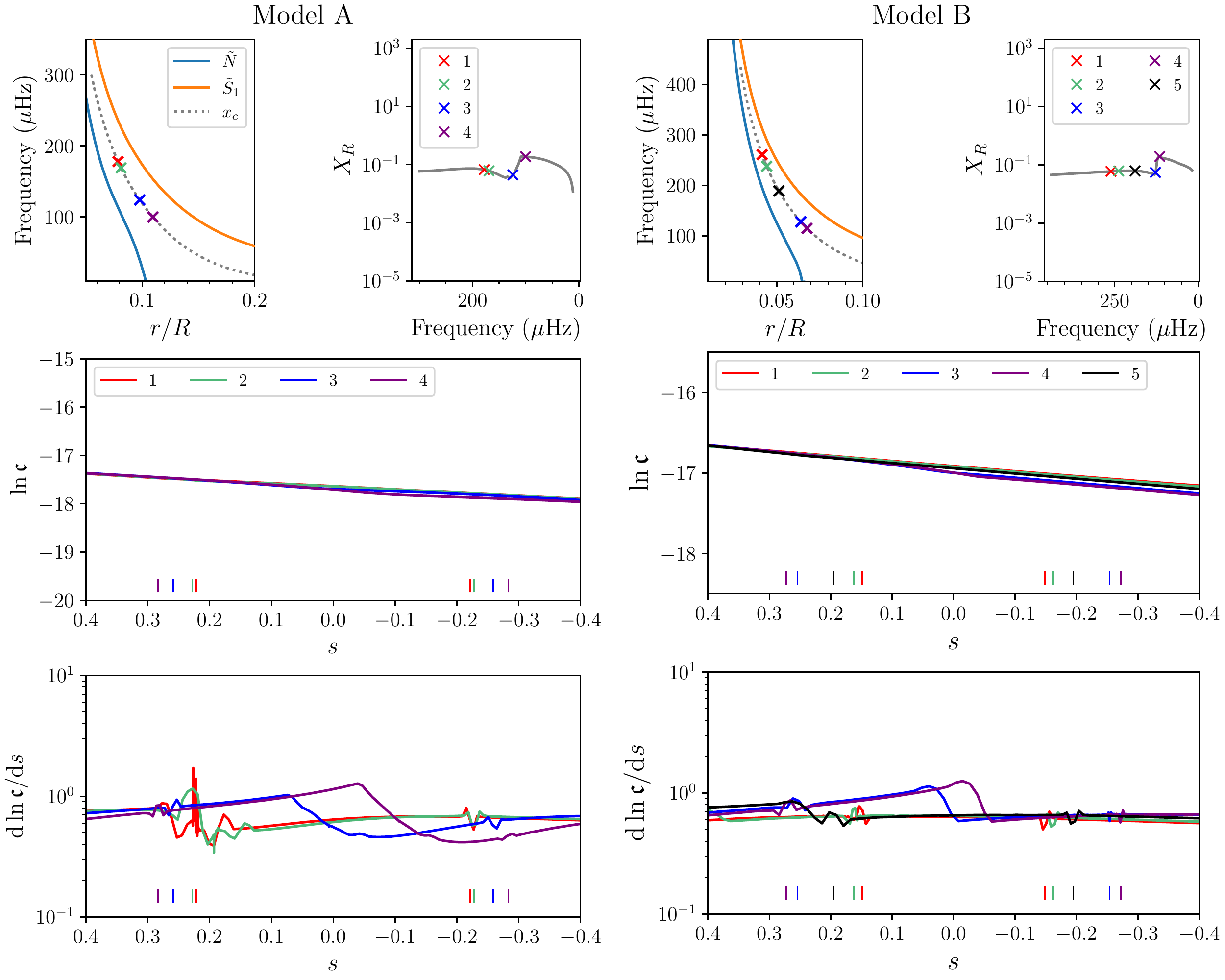}}
\caption{$\ln \mathfrak{c}$ as a function of $s$ for modes with the same frequencies as in Figure~\ref{fg:lnc_ori}. The changes in the Lamb and buoyancy frequencies induced by the structural variations are artificially removed in both Model A and Model B so that $\ln \mathfrak{c}$ are only perturbed by the gradient change of the characteristic frequencies near the BCZ. The bottom row panels show $\mathrm d \ln \mathfrak{c} / \mathrm d s$ to highlight the variations near the center of the EVZ ($s=0$). The turning points $s_0$ ($x_1$) and $-s_0$ ($x_2$), locations of which are indicated by the colored vertical bars in the lower four panels, are singularities of $\mathfrak{c}$. As a result, extra fluctuations are present in $\mathrm d \ln \mathfrak{c} / \mathrm d s$ (and also in $\ln \mathfrak{c}$, but invisible in the middle panels) around the turning points.}
\label{fg:lnc_bcz}
\end{figure*}

\subsection{Contribution of Buoyancy Spike}
\label{sc:ng}

Figure~\ref{fg:lnc_ng} illustrates the cases in which the displacement in $\tilde S_1$ is artificially removed from the stellar models, allowing for the study of the contribution of the buoyancy spike (glitch). The two large-scale jumps caused by the displacement in $\tilde S_1$ at $s_{\rm spike}$ are also eliminated for Mode 1 and Mode 2. Thus, the significant variations in $\ln \mathfrak{c}$ are mainly induced by the buoyancy spike (glitch), which can be proved by the fact that the locations of these variations overlap with $s_{\rm spike}$ for each mode (see Figure~\ref{fg:lnc_ng}). In this case, the buoyancy spike (glitch) gives rise to stronger variations in $X_R$ than the Lamb frequency displacement, which can be seen from the comparisons between Figure~\ref{fg:lnc_s1} and Figure~\ref{fg:lnc_ng} for Mode 3 in Model A and Mode 5 in Model B. 

For Model A, the sharp variations of all four modes are generally above the straight lines representing the unperturbed descending trend of $\ln \mathfrak{c}$. When the spike is near the centre of the EVZ, as in the case of Mode 3, dramatic variations in $X_R$ are resulted. However, tiny differences in terms of the descending slopes are also seen in $\ln \mathfrak{c}$ between positions above and below each spike, which cause further deviations in $\ln \mathfrak{c}$ at $s=0$ and contribute to the sudden increase in $X_R$ of Mode 4 compared the other three modes. In fact, the sudden increase occurs as soon as the spike is deeper than the center of the EVZ for a certain mode. 

For Model B, the glitch impacts the modes in a similar way by generating large-scale variations at $s_{\rm spike}$. Nevertheless, the protruding structures of $\ln \mathfrak{c}$ of Mode 3 and Mode 4 are now below the straight lines. This is because $Q$ changes its sign at the inner turning point $-s_0$. An illustration of such situations is given in Figure~\ref{fg:comp_q} in which the buoyancy glitch (the sharp signature in $|Q|$) is deeper than the inner turning point (the sharp signature in $|s+s_0|$) and outside the EVZ for Mode 3 but it is close to the centre of the EVZ for Mode 5. The values of $Q$ within the area of the glitch is increased for Mode 3, while they are reduced for Mode 5. According to equation~\eqref{eq:c}, $\mathfrak{c}$ is inversely proportional to $Q$, thus the sharp signatures change pointing direction. Note in this case, $P$ is spikeless so the main contributing factor to the sharp variations is the buoyancy spike (glitch), and $Q$ varies nearly in parallel with $s+s_0$, leading to the straight lines of $\ln \mathfrak{c}$ in regions not affected by the spike (glitch). But in the vicinity of the spike (glitch), gradient of $\ln \mathfrak{c}$ differs between the sides of the spike (glitch) which leading to the increase in $X_R$ for Mode 4.

\subsection{Contribution of Gradient Change of Characteristic Frequencies}
\label{sc:bcz}

With both the $\tilde S_1$ displacement and the buoyancy spike (glitch) removed from the models, the only factor left to perturb the value of $X_R$ and asymptotic $q$ is the gradient change of the characteristic frequencies near the BCZ. As shown in Figure~\ref{fg:lnc_bcz} for both Model A and Model B that are free of sharp changes in the characteristic frequencies, $\ln \mathfrak{c}$ behaves almost linearly with respect to $s$ in the range encompassing the EVZ. However, subtle deviations in $\ln \mathfrak{c}$ are observable in negative $s$ region for Mode 3 and Mode 4 in both models, which are highlighted by the significant drop in $\mathrm d \ln \mathfrak{c} / \mathrm d s$ values near $s=0$ shown in the bottom row panels in Figure~\ref{fg:lnc_bcz}. The deviations are due to the fact that $\tilde N$ and $\tilde S_1$ vary differently with radius, and hence have different derivatives with respect to $s$ between the layers below and above the BCZ, which propagate to $X_R$ through equations~\eqref{eq:xr} and~\eqref{eq:c}. $\tilde N$ vary nearly in parallel with $\tilde S_1$ in regions below the BCZ, but $\tilde N^2$ becomes negative in the convective zone that is above the BCZ while $\tilde S_1$ keeps positive. This variation due to gradient change of the characteristic frequencies at the BCZ is the reason for the jump in $X_R$ between Mode 3 and Mode 4 when both the displacement in $\tilde S_1$ and the buoyancy spike (glitch) are removed. The detection of the impact of the gradient change of the characteristic frequencies at the BCZ requires the EVZ centre is close to the BCZ, i.e., in evolved RGB stars where $\numax$ are low enough. And the difference of variations between $\tilde N$ and $\tilde S_1$ with radius becomes larger when stars get even older, leading to a faster drop in $q$. Note the fluctuations in $\mathrm d \ln \mathfrak{c} / \mathrm d s$ near the turning points are numerically caused by the fact $s_0$ and $-s_0$ are singularities of $\mathfrak{c}$.

Unlike the the displacement in $\tilde S_1$ and the buoyancy spike (glitch) that are factors not considered in the asymptotic analysis, the gradient change of the characteristic frequencies at the BCZ  should have an effect on the asymptotic $q$ even when the first two factors are not present. Thus, the contribution of the gradient change of the characteristic frequencies should also be reflected by the observations. Comprehensive studies of the evolution of mixed-mode properties with age using stellar models or using observational data of large sample of evolved RGB stars will help us better understand how the gradient change of the characteristic frequencies at the BCZ impacts the oscillations, but it is beyond the scope of this paper.

\section{Conclusion}
\label{sc:conclusion}

In this paper, the impact of the structural variations on the mixed-mode coupling factor, $q$, computed using  the asymptotic formalism from \cite{takata2016} are addressed from a theoretical point of view. The structural variations located in the mid-layers of red giant stars are caused by the chemical discontinuity left behind during the first dredge-up, which produce sharp variations in the characteristic frequencies. Since the asymptotic analysis assumes smooth variation of the characteristic frequencies with radius, one would expect the inferred asymptotic $q$ to behave abnormally with the presence of the structural variations. By utilizing stellar models on the red giant branch showing significant structural variations in a region encompassing the evanescent zone, three main contributing factors, associated either with the structural variations or the gradient change of the characteristic frequencies, to the variations in asymptotic $q$ are discussed in detail.  

First, the structural variations induce a displacement in the Lamb frequency that increases the value of asymptotic $q$ when the displacement starts to be located inside the evanescent zone. This happens in relatively low-luminosity red giants in which the $\numax$ is high and the detectable oscillation modes with frequencies around $\numax$ have a thin evanescent zone. Second, the structural variations lead to sharp changes in the buoyancy frequency, which is called the buoyancy spike in low-luminosity red giants, or the buoyancy glitch in the case of evolved red giants. The buoyancy spike (glitch) brings in dramatic variations in asymptotic $q$ when it is positioned near the centre of the evanescent zone. Indeed, it has been demonstrated with stellar models, modified by explicitly removing either the Lamb frequency displacement or the buoyancy spike (glitch), that the largest contributions to the variations in $X_R$, hence the asymptotic $q$, come from the buoyancy spike (glitch).
Since in red giants the buoyancy spike (glitch) emerges near the outer edge of the g-mode cavity coinciding with the base of the convective zone, the impact of the buoyancy spike (glitch) can be felt by detectable modes that have an intermediate-size evanescent zone.
Furthermore, due to the fact the asymptotic analysis of mixed modes assumes smooth background free of structural variations, the asymptotic $q$ is most affected in these modes possessing an intermediate-size evanescent zone. Thus, a new formalism is needed to account for the impact of structural variations. 
Actually, the rapid structural variation can be assimilated to a barrier where oscillation modes are partially reflected and transmitted, just like in the evanescent zone. Thus, the spike (glitch) can be again characterized in the framework of \cite{takata2016}, by a reflection coefficient and a phase lag at the reflection. Promising progresses have been presented in recent works of \cite{pincon2018} as well as in  \cite{pincon2022}.

Third, the asymptotic $q$ varies smoothly with mode frequency if the model is free of structural variations. However, the gradient change of the characteristic frequencies, corresponding to their different variations with radius near the base of the convective zone, also changes the value of $q$ when the centre of the evanescent zone of the mode with a certain frequency is in the vicinity of the base of the convective zone. As a result, a sudden increase in asymptotic $q$ occurs, followed by a quick drop when the evanescent zone is well above the base of the convective zone. This effect of the gradient change of the characteristic frequencies at the base of the convective zone can be reflected by observations of evolved red giants in which the low value of $\numax$ ensures that the detectable oscillation modes have a thick evanescent zone located above the base of the convective zone.

This work will have an influence on our understanding of the internal structure of stars and how it evolves with age.
The signature of the structural variations on the oscillation frequencies is rather small so that highly precise frequencies are required for a successful detection. Focusing on a large number of bright stars with several years of ultrahigh quality photometry observations, the upcoming space missions like ESA's PLATO and China's Earth 2.0 will have the potential to provide detections of minor frequency changes due to the structural variations.

%\backmatter

\section*{Acknowledgments}

The author gratefully thanks J\o{}rgen Christensen-Dalsgaard and Margarida Cunha for the discussions on the analysis.

%\nocite{*}% Show all bib entries - both cited and uncited; comment this line to view only cited bib entries;
\bibliography{draft}%

\begin{thebibliography}{}

\bibitem [\protect \citeauthoryear {%
{Aerts}%
, {Christensen-Dalsgaard}%
\BCBL {}\ \BBA {} {Kurtz}%
}{%
{Aerts}%
\ \protect \BOthers {.}}{%
{\protect \APACyear {2010}}%
}]{%
aerts2010}
\APACinsertmetastar {%
aerts2010}%
\begin{APACrefauthors}%
{Aerts}, C.%
, {Christensen-Dalsgaard}, J.%
\BCBL {}\ \BBA {} {Kurtz}, D\BPBI W.%
\end{APACrefauthors}%
\unskip\
\newblock
\APACrefYear{2010},
\newblock
\APACrefbtitle {{Asteroseismology}} {{Asteroseismology}}.
\newblock
\APACaddressPublisher{}{Springer Netherlands}.
\PrintBackRefs{\CurrentBib}

\bibitem [\protect \citeauthoryear {%
{Baglin}%
\ \protect \BOthers {.}}{%
{Baglin}%
\ \protect \BOthers {.}}{%
{\protect \APACyear {2006}}%
}]{%
baglin2006}
\APACinsertmetastar {%
baglin2006}%
\begin{APACrefauthors}%
{Baglin}, A.%
, {Auvergne}, M.%
, {Barge}, P.%
\ et al.\end{APACrefauthors}%
\unskip\
\newblock
\APACrefYearMonthDay{2006}{{\APACmonth{11}}}{},
\newblock
{\BBOQ}\APACrefatitle {{Scientific Objectives for a Minisat: CoRoT}}
  {{Scientific Objectives for a Minisat: CoRoT}}.{\BBCQ}
\newblock
\BIn{} M.~{Fridlund}, A.~{Baglin}, J.~{Lochard}\BCBL {}\ \BOthers {.}\ (\BEDS),
  \APACrefbtitle {The CoRoT Mission Pre-Launch Status - Stellar Seismology and
  Planet Finding} {The CoRoT Mission Pre-Launch Status - Stellar Seismology and
  Planet Finding}\ \BVOL\ 1306, \BPG~33.
\PrintBackRefs{\CurrentBib}

\bibitem [\protect \citeauthoryear {%
{Beck}%
\ \protect \BOthers {.}}{%
{Beck}%
\ \protect \BOthers {.}}{%
{\protect \APACyear {2012}}%
}]{%
beck2012}
\APACinsertmetastar {%
beck2012}%
\begin{APACrefauthors}%
{Beck}, P\BPBI G.%
, {Montalban}, J.%
, {Kallinger}, T.%
\ et al.\end{APACrefauthors}%
\unskip\
\newblock
\APACrefYearMonthDay{2012}{{\APACmonth{01}}}{},
\newblock
\unskip
\newblock
\APACjournalVolNumPages{\nat}{481}{7379}{55-57}.
\newblock
\begin{APACrefDOI} \doi{10.1038/nature10612} \end{APACrefDOI}
\PrintBackRefs{\CurrentBib}

\bibitem [\protect \citeauthoryear {%
{Bedding}%
\ \protect \BOthers {.}}{%
{Bedding}%
\ \protect \BOthers {.}}{%
{\protect \APACyear {2007}}%
}]{%
bedding2007}
\APACinsertmetastar {%
bedding2007}%
\begin{APACrefauthors}%
{Bedding}, T\BPBI R.%
, {Kjeldsen}, H.%
, {Arentoft}, T.%
\ et al.\end{APACrefauthors}%
\unskip\
\newblock
\APACrefYearMonthDay{2007}{{\APACmonth{07}}}{},
\newblock
\unskip
\newblock
\APACjournalVolNumPages{\apj}{663}{2}{1315-1324}.
\newblock
\begin{APACrefDOI} \doi{10.1086/518593} \end{APACrefDOI}
\PrintBackRefs{\CurrentBib}

\bibitem [\protect \citeauthoryear {%
{Bedding}%
\ \protect \BOthers {.}}{%
{Bedding}%
\ \protect \BOthers {.}}{%
{\protect \APACyear {2011}}%
}]{%
bedding2011}
\APACinsertmetastar {%
bedding2011}%
\begin{APACrefauthors}%
{Bedding}, T\BPBI R.%
, {Mosser}, B.%
, {Huber}, D.%
\ et al.\end{APACrefauthors}%
\unskip\
\newblock
\APACrefYearMonthDay{2011}{{\APACmonth{03}}}{},
\newblock
\unskip
\newblock
\APACjournalVolNumPages{\nat}{471}{7340}{608-611}.
\newblock
\begin{APACrefDOI} \doi{10.1038/nature09935} \end{APACrefDOI}
\PrintBackRefs{\CurrentBib}

\bibitem [\protect \citeauthoryear {%
{B{\"o}hm-Vitense}%
}{%
{B{\"o}hm-Vitense}%
}{%
{\protect \APACyear {1958}}%
}]{%
bohm1958}
\APACinsertmetastar {%
bohm1958}%
\begin{APACrefauthors}%
{B{\"o}hm-Vitense}, E.%
\end{APACrefauthors}%
\unskip\
\newblock
\APACrefYearMonthDay{1958}{{\APACmonth{01}}}{},
\newblock
\unskip
\newblock
\APACjournalVolNumPages{\zap}{46}{}{108}.
\PrintBackRefs{\CurrentBib}

\bibitem [\protect \citeauthoryear {%
{Borucki}%
\ \protect \BOthers {.}}{%
{Borucki}%
\ \protect \BOthers {.}}{%
{\protect \APACyear {2010}}%
}]{%
borucki2010}
\APACinsertmetastar {%
borucki2010}%
\begin{APACrefauthors}%
{Borucki}, W\BPBI J.%
, {Koch}, D.%
, {Basri}, G.%
\ et al.\end{APACrefauthors}%
\unskip\
\newblock
\APACrefYearMonthDay{2010}{Feb}{},
\newblock
\unskip
\newblock
\APACjournalVolNumPages{Science}{327}{5968}{977}.
\newblock
\begin{APACrefDOI} \doi{10.1126/science.1185402} \end{APACrefDOI}
\PrintBackRefs{\CurrentBib}

\bibitem [\protect \citeauthoryear {%
{Carrier}%
, {Eggenberger}%
\BCBL {}\ \BBA {} {Bouchy}%
}{%
{Carrier}%
\ \protect \BOthers {.}}{%
{\protect \APACyear {2005}}%
}]{%
carrier2005}
\APACinsertmetastar {%
carrier2005}%
\begin{APACrefauthors}%
{Carrier}, F.%
, {Eggenberger}, P.%
\BCBL {}\ \BBA {} {Bouchy}, F.%
\end{APACrefauthors}%
\unskip\
\newblock
\APACrefYearMonthDay{2005}{{\APACmonth{05}}}{},
\newblock
\unskip
\newblock
\APACjournalVolNumPages{\aap}{434}{3}{1085-1095}.
\newblock
\begin{APACrefDOI} \doi{10.1051/0004-6361:20042140} \end{APACrefDOI}
\PrintBackRefs{\CurrentBib}

\bibitem [\protect \citeauthoryear {%
{Christensen-Dalsgaard}%
}{%
{Christensen-Dalsgaard}%
}{%
{\protect \APACyear {2008}}%
{\protect \APACexlab {{\protect \BCnt {1}}}}}]{%
jcd08b}
\APACinsertmetastar {%
jcd08b}%
\begin{APACrefauthors}%
{Christensen-Dalsgaard}, J.%
\end{APACrefauthors}%
\unskip\
\newblock
\APACrefYearMonthDay{2008{\protect \BCnt {1}}}{{\APACmonth{08}}}{},
\newblock
\unskip
\newblock
\APACjournalVolNumPages{\apss}{316}{1-4}{113-120}.
\newblock
\begin{APACrefDOI} \doi{10.1007/s10509-007-9689-z} \end{APACrefDOI}
\PrintBackRefs{\CurrentBib}

\bibitem [\protect \citeauthoryear {%
{Christensen-Dalsgaard}%
}{%
{Christensen-Dalsgaard}%
}{%
{\protect \APACyear {2008}}%
{\protect \APACexlab {{\protect \BCnt {2}}}}}]{%
jcd08a}
\APACinsertmetastar {%
jcd08a}%
\begin{APACrefauthors}%
{Christensen-Dalsgaard}, J.%
\end{APACrefauthors}%
\unskip\
\newblock
\APACrefYearMonthDay{2008{\protect \BCnt {2}}}{{\APACmonth{08}}}{},
\newblock
\unskip
\newblock
\APACjournalVolNumPages{\apss}{316}{1-4}{13-24}.
\newblock
\begin{APACrefDOI} \doi{10.1007/s10509-007-9675-5} \end{APACrefDOI}
\PrintBackRefs{\CurrentBib}

\bibitem [\protect \citeauthoryear {%
{Christensen-Dalsgaard}%
}{%
{Christensen-Dalsgaard}%
}{%
{\protect \APACyear {2015}}%
}]{%
jcd2015}
\APACinsertmetastar {%
jcd2015}%
\begin{APACrefauthors}%
{Christensen-Dalsgaard}, J.%
\end{APACrefauthors}%
\unskip\
\newblock
\APACrefYearMonthDay{2015}{{\APACmonth{10}}}{},
\newblock
\unskip
\newblock
\APACjournalVolNumPages{\mnras}{453}{1}{666-670}.
\newblock
\begin{APACrefDOI} \doi{10.1093/mnras/stv1656} \end{APACrefDOI}
\PrintBackRefs{\CurrentBib}

\bibitem [\protect \citeauthoryear {%
{Cowling}%
}{%
{Cowling}%
}{%
{\protect \APACyear {1941}}%
}]{%
cowling}
\APACinsertmetastar {%
cowling}%
\begin{APACrefauthors}%
{Cowling}, T\BPBI G.%
\end{APACrefauthors}%
\unskip\
\newblock
\APACrefYearMonthDay{1941}{{\APACmonth{01}}}{},
\newblock
\unskip
\newblock
\APACjournalVolNumPages{\mnras}{101}{}{367}.
\newblock
\begin{APACrefDOI} \doi{10.1093/mnras/101.8.367} \end{APACrefDOI}
\PrintBackRefs{\CurrentBib}

\bibitem [\protect \citeauthoryear {%
{Cunha}%
\ \protect \BOthers {.}}{%
{Cunha}%
\ \protect \BOthers {.}}{%
{\protect \APACyear {2019}}%
}]{%
cunha2019}
\APACinsertmetastar {%
cunha2019}%
\begin{APACrefauthors}%
{Cunha}, M\BPBI S.%
, {Avelino}, P\BPBI P.%
, {Christensen-Dalsgaard}, J.%
, {Stello}, D.%
, {Vrard}, M.%
, {Jiang}, C.%
\BCBL {}\ \BBA {} {Mosser}, B.%
\end{APACrefauthors}%
\unskip\
\newblock
\APACrefYearMonthDay{2019}{{\APACmonth{11}}}{},
\newblock
\unskip
\newblock
\APACjournalVolNumPages{\mnras}{490}{1}{909-926}.
\newblock
\begin{APACrefDOI} \doi{10.1093/mnras/stz2582} \end{APACrefDOI}
\PrintBackRefs{\CurrentBib}

\bibitem [\protect \citeauthoryear {%
{Cunha}%
, {Stello}%
, {Avelino}%
, {Christensen-Dalsgaard}%
\BCBL {}\ \BBA {} {Townsend}%
}{%
{Cunha}%
\ \protect \BOthers {.}}{%
{\protect \APACyear {2015}}%
}]{%
cunha2015}
\APACinsertmetastar {%
cunha2015}%
\begin{APACrefauthors}%
{Cunha}, M\BPBI S.%
, {Stello}, D.%
, {Avelino}, P\BPBI P.%
, {Christensen-Dalsgaard}, J.%
\BCBL {}\ \BBA {} {Townsend}, R\BPBI H\BPBI D.%
\end{APACrefauthors}%
\unskip\
\newblock
\APACrefYearMonthDay{2015}{{\APACmonth{06}}}{},
\newblock
\unskip
\newblock
\APACjournalVolNumPages{\apj}{805}{2}{127}.
\newblock
\begin{APACrefDOI} \doi{10.1088/0004-637X/805/2/127} \end{APACrefDOI}
\PrintBackRefs{\CurrentBib}

\bibitem [\protect \citeauthoryear {%
{Gizon}%
\ \protect \BOthers {.}}{%
{Gizon}%
\ \protect \BOthers {.}}{%
{\protect \APACyear {2013}}%
}]{%
gizon2013}
\APACinsertmetastar {%
gizon2013}%
\begin{APACrefauthors}%
{Gizon}, L.%
, {Ballot}, J.%
, {Michel}, E.%
\ et al.\end{APACrefauthors}%
\unskip\
\newblock
\APACrefYearMonthDay{2013}{{\APACmonth{08}}}{},
\newblock
\unskip
\newblock
\APACjournalVolNumPages{Proceedings of the National Academy of
  Science}{110}{33}{13267-13271}.
\newblock
\begin{APACrefDOI} \doi{10.1073/pnas.1303291110} \end{APACrefDOI}
\PrintBackRefs{\CurrentBib}

\bibitem [\protect \citeauthoryear {%
{Hekker}%
\ \BBA {} {Christensen-Dalsgaard}%
}{%
{Hekker}%
\ \BBA {} {Christensen-Dalsgaard}%
}{%
{\protect \APACyear {2017}}%
}]{%
hekker2017}
\APACinsertmetastar {%
hekker2017}%
\begin{APACrefauthors}%
{Hekker}, S.%
\BCBT {}\ \BBA {} {Christensen-Dalsgaard}, J.%
\end{APACrefauthors}%
\unskip\
\newblock
\APACrefYearMonthDay{2017}{{\APACmonth{06}}}{},
\newblock
\unskip
\newblock
\APACjournalVolNumPages{\aapr}{25}{1}{1}.
\newblock
\begin{APACrefDOI} \doi{10.1007/s00159-017-0101-x} \end{APACrefDOI}
\PrintBackRefs{\CurrentBib}

\bibitem [\protect \citeauthoryear {%
{Huber}%
\ \protect \BOthers {.}}{%
{Huber}%
\ \protect \BOthers {.}}{%
{\protect \APACyear {2010}}%
}]{%
huber2010}
\APACinsertmetastar {%
huber2010}%
\begin{APACrefauthors}%
{Huber}, D.%
, {Bedding}, T\BPBI R.%
, {Stello}, D.%
\ et al.\end{APACrefauthors}%
\unskip\
\newblock
\APACrefYearMonthDay{2010}{{\APACmonth{11}}}{},
\newblock
\unskip
\newblock
\APACjournalVolNumPages{\apj}{723}{2}{1607-1617}.
\newblock
\begin{APACrefDOI} \doi{10.1088/0004-637X/723/2/1607} \end{APACrefDOI}
\PrintBackRefs{\CurrentBib}

\bibitem [\protect \citeauthoryear {%
{Jiang}%
, {Cunha}%
, {Christensen-Dalsgaard}%
\BCBL {}\ \BBA {} {Zhang}%
}{%
{Jiang}%
\ \protect \BOthers {.}}{%
{\protect \APACyear {2020}}%
}]{%
jiang2020}
\APACinsertmetastar {%
jiang2020}%
\begin{APACrefauthors}%
{Jiang}, C.%
, {Cunha}, M.%
, {Christensen-Dalsgaard}, J.%
\BCBL {}\ \BBA {} {Zhang}, Q.%
\end{APACrefauthors}%
\unskip\
\newblock
\APACrefYearMonthDay{2020}{{\APACmonth{06}}}{},
\newblock
\unskip
\newblock
\APACjournalVolNumPages{\mnras}{495}{1}{621-636}.
\newblock
\begin{APACrefDOI} \doi{10.1093/mnras/staa1285} \end{APACrefDOI}
\PrintBackRefs{\CurrentBib}

\bibitem [\protect \citeauthoryear {%
{Jiang}%
, {Cunha}%
, {Christensen-Dalsgaard}%
, {Zhang}%
\BCBL {}\ \BBA {} {Gizon}%
}{%
{Jiang}%
\ \protect \BOthers {.}}{%
{\protect \APACyear {2022}}%
}]{%
jiang2022}
\APACinsertmetastar {%
jiang2022}%
\begin{APACrefauthors}%
{Jiang}, C.%
, {Cunha}, M.%
, {Christensen-Dalsgaard}, J.%
, {Zhang}, Q\BPBI S.%
\BCBL {}\ \BBA {} {Gizon}, L.%
\end{APACrefauthors}%
\unskip\
\newblock
\APACrefYearMonthDay{2022}{{\APACmonth{09}}}{},
\newblock
\unskip
\newblock
\APACjournalVolNumPages{\mnras}{515}{3}{3853-3866}.
\newblock
\begin{APACrefDOI} \doi{10.1093/mnras/stac2065} \end{APACrefDOI}
\PrintBackRefs{\CurrentBib}

\bibitem [\protect \citeauthoryear {%
{Jiang}%
\ \protect \BOthers {.}}{%
{Jiang}%
\ \protect \BOthers {.}}{%
{\protect \APACyear {2011}}%
}]{%
jiang2011}
\APACinsertmetastar {%
jiang2011}%
\begin{APACrefauthors}%
{Jiang}, C.%
, {Jiang}, B\BPBI W.%
, {Christensen-Dalsgaard}, J.%
\ et al.\end{APACrefauthors}%
\unskip\
\newblock
\APACrefYearMonthDay{2011}{{\APACmonth{12}}}{},
\newblock
\unskip
\newblock
\APACjournalVolNumPages{\apj}{742}{2}{120}.
\newblock
\begin{APACrefDOI} \doi{10.1088/0004-637X/742/2/120} \end{APACrefDOI}
\PrintBackRefs{\CurrentBib}

\bibitem [\protect \citeauthoryear {%
{Kallinger}%
\ \protect \BOthers {.}}{%
{Kallinger}%
\ \protect \BOthers {.}}{%
{\protect \APACyear {2012}}%
}]{%
kallinger2012}
\APACinsertmetastar {%
kallinger2012}%
\begin{APACrefauthors}%
{Kallinger}, T.%
, {Hekker}, S.%
, {Mosser}, B.%
\ et al.\end{APACrefauthors}%
\unskip\
\newblock
\APACrefYearMonthDay{2012}{{\APACmonth{05}}}{},
\newblock
\unskip
\newblock
\APACjournalVolNumPages{\aap}{541}{}{A51}.
\newblock
\begin{APACrefDOI} \doi{10.1051/0004-6361/201218854} \end{APACrefDOI}
\PrintBackRefs{\CurrentBib}

\bibitem [\protect \citeauthoryear {%
{Mosser}%
\ \protect \BOthers {.}}{%
{Mosser}%
\ \protect \BOthers {.}}{%
{\protect \APACyear {2012}}%
}]{%
mosser2012}
\APACinsertmetastar {%
mosser2012}%
\begin{APACrefauthors}%
{Mosser}, B.%
, {Goupil}, M\BPBI J.%
, {Belkacem}, K.%
\ et al.\end{APACrefauthors}%
\unskip\
\newblock
\APACrefYearMonthDay{2012}{{\APACmonth{04}}}{},
\newblock
\unskip
\newblock
\APACjournalVolNumPages{\aap}{540}{}{A143}.
\newblock
\begin{APACrefDOI} \doi{10.1051/0004-6361/201118519} \end{APACrefDOI}
\PrintBackRefs{\CurrentBib}

\bibitem [\protect \citeauthoryear {%
{Mosser}%
, {Miglio}%
\BCBL {}\ \BBA {} {CoRot Team}%
}{%
{Mosser}%
\ \protect \BOthers {.}}{%
{\protect \APACyear {2016}}%
}]{%
mosser2016}
\APACinsertmetastar {%
mosser2016}%
\begin{APACrefauthors}%
{Mosser}, B.%
, {Miglio}, A.%
\BCBL {}\ \BBA {} {CoRot Team}.%
\end{APACrefauthors}%
\unskip\
\newblock
\APACrefYearMonthDay{2016}{}{},
\newblock
{\BBOQ}\APACrefatitle {{IV.2 Pulsating red giant stars}} {{IV.2 Pulsating red
  giant stars}}.{\BBCQ}
\newblock
\BIn{} \APACrefbtitle {The CoRoT Legacy Book: The Adventure of the Ultra High
  Precision Photometry from Space} {The CoRoT Legacy Book: The Adventure of the
  Ultra High Precision Photometry from Space}\ \BPG~197.
\newblock
\begin{APACrefDOI} \doi{10.1051/978-2-7598-1876-1.c042} \end{APACrefDOI}
\PrintBackRefs{\CurrentBib}

\bibitem [\protect \citeauthoryear {%
{Pin{\c{c}}on}%
}{%
{Pin{\c{c}}on}%
}{%
{\protect \APACyear {2018}}%
}]{%
pincon2018}
\APACinsertmetastar {%
pincon2018}%
\begin{APACrefauthors}%
{Pin{\c{c}}on}, C.%
\end{APACrefauthors}%
\unskip\
\newblock
\APACrefYearMonthDay{2018}{{\APACmonth{09}}}{},
\newblock
{\BBOQ}\APACrefatitle {{A simple representation of oscillation modes in stars:
  from mixed modes coupling to glitches}} {{A simple representation of
  oscillation modes in stars: from mixed modes coupling to glitches}}.{\BBCQ}
\newblock
\BIn{} \APACrefbtitle {PHysics of Oscillating STars. Proceedings from the PHOST
  (PHysics of Oscillating STars) symposium hosted by the Oceanographic
  Observatory in Banyuls-sur-mer (France) from 2-7 September 2018. This
  conference honours the life work of Professor Hiromoto Shibahashi} {PHysics
  of Oscillating STars. Proceedings from the PHOST (PHysics of Oscillating
  STars) symposium hosted by the Oceanographic Observatory in Banyuls-sur-mer
  (France) from 2-7 September 2018. This conference honours the life work of
  Professor Hiromoto Shibahashi}\ \BPG~46.
\newblock
\begin{APACrefDOI} \doi{10.5281/zenodo.2207751} \end{APACrefDOI}
\PrintBackRefs{\CurrentBib}

\bibitem [\protect \citeauthoryear {%
{Pin{\c{c}}on}%
, {Goupil}%
\BCBL {}\ \BBA {} {Belkacem}%
}{%
{Pin{\c{c}}on}%
\ \protect \BOthers {.}}{%
{\protect \APACyear {2020}}%
}]{%
pincon2020}
\APACinsertmetastar {%
pincon2020}%
\begin{APACrefauthors}%
{Pin{\c{c}}on}, C.%
, {Goupil}, M\BPBI J.%
\BCBL {}\ \BBA {} {Belkacem}, K.%
\end{APACrefauthors}%
\unskip\
\newblock
\APACrefYearMonthDay{2020}{{\APACmonth{02}}}{},
\newblock
\unskip
\newblock
\APACjournalVolNumPages{\aap}{634}{}{A68}.
\newblock
\begin{APACrefDOI} \doi{10.1051/0004-6361/201936864} \end{APACrefDOI}
\PrintBackRefs{\CurrentBib}

\bibitem [\protect \citeauthoryear {%
{Pin{\c{c}}on}%
\ \BBA {} {Takata}%
}{%
{Pin{\c{c}}on}%
\ \BBA {} {Takata}%
}{%
{\protect \APACyear {2022}}%
}]{%
pincon2022}
\APACinsertmetastar {%
pincon2022}%
\begin{APACrefauthors}%
{Pin{\c{c}}on}, C.%
\BCBT {}\ \BBA {} {Takata}, M.%
\end{APACrefauthors}%
\unskip\
\newblock
\APACrefYearMonthDay{2022}{{\APACmonth{05}}}{},
\newblock
\unskip
\newblock
\APACjournalVolNumPages{\aap}{661}{}{A139}.
\newblock
\begin{APACrefDOI} \doi{10.1051/0004-6361/202243157} \end{APACrefDOI}
\PrintBackRefs{\CurrentBib}

\bibitem [\protect \citeauthoryear {%
{Pin{\c{c}}on}%
, {Takata}%
\BCBL {}\ \BBA {} {Mosser}%
}{%
{Pin{\c{c}}on}%
\ \protect \BOthers {.}}{%
{\protect \APACyear {2019}}%
}]{%
pincon2019}
\APACinsertmetastar {%
pincon2019}%
\begin{APACrefauthors}%
{Pin{\c{c}}on}, C.%
, {Takata}, M.%
\BCBL {}\ \BBA {} {Mosser}, B.%
\end{APACrefauthors}%
\unskip\
\newblock
\APACrefYearMonthDay{2019}{{\APACmonth{06}}}{},
\newblock
\unskip
\newblock
\APACjournalVolNumPages{\aap}{626}{}{A125}.
\newblock
\begin{APACrefDOI} \doi{10.1051/0004-6361/201935327} \end{APACrefDOI}
\PrintBackRefs{\CurrentBib}

\bibitem [\protect \citeauthoryear {%
{Ricker}%
\ \protect \BOthers {.}}{%
{Ricker}%
\ \protect \BOthers {.}}{%
{\protect \APACyear {2014}}%
}]{%
ricker2014}
\APACinsertmetastar {%
ricker2014}%
\begin{APACrefauthors}%
{Ricker}, G\BPBI R.%
, {Winn}, J\BPBI N.%
, {Vanderspek}, R.%
\ et al.\end{APACrefauthors}%
\unskip\
\newblock
\APACrefYearMonthDay{2014}{}{},
\newblock
{\BBOQ}\APACrefatitle {{Transiting Exoplanet Survey Satellite (TESS)}}
  {{Transiting Exoplanet Survey Satellite (TESS)}}.{\BBCQ}
\newblock
\BIn{} \APACrefbtitle {Proceedings of the SPIE, Volume 9143, id. 914320 15 pp.
  (2014).} {Proceedings of the SPIE, Volume 9143, id. 914320 15 pp. (2014).}\
  \BVOL\ 9143, \BPG~914320.
\newblock
\begin{APACrefDOI} \doi{10.1117/12.2063489} \end{APACrefDOI}
\PrintBackRefs{\CurrentBib}

\bibitem [\protect \citeauthoryear {%
{Shibahashi}%
}{%
{Shibahashi}%
}{%
{\protect \APACyear {1979}}%
}]{%
shibahashi1979}
\APACinsertmetastar {%
shibahashi1979}%
\begin{APACrefauthors}%
{Shibahashi}, H.%
\end{APACrefauthors}%
\unskip\
\newblock
\APACrefYearMonthDay{1979}{{\APACmonth{01}}}{},
\newblock
\unskip
\newblock
\APACjournalVolNumPages{\pasj}{31}{}{87-104}.
\PrintBackRefs{\CurrentBib}

\bibitem [\protect \citeauthoryear {%
{Takata}%
}{%
{Takata}%
}{%
{\protect \APACyear {2006}}%
}]{%
takata2006}
\APACinsertmetastar {%
takata2006}%
\begin{APACrefauthors}%
{Takata}, M.%
\end{APACrefauthors}%
\unskip\
\newblock
\APACrefYearMonthDay{2006}{{\APACmonth{10}}}{},
\newblock
{\BBOQ}\APACrefatitle {{Rigorous analysis of dipolar oscillations of stars}}
  {{Rigorous analysis of dipolar oscillations of stars}}.{\BBCQ}
\newblock
\BIn{} K.~{Fletcher}\ \BBA {} M.~{Thompson}\ (\BEDS), \APACrefbtitle
  {Proceedings of SOHO 18/GONG 2006/HELAS I, Beyond the spherical Sun}
  {Proceedings of SOHO 18/GONG 2006/HELAS I, Beyond the spherical Sun}\
  \BVOL~624, \BPG~26.
\PrintBackRefs{\CurrentBib}

\bibitem [\protect \citeauthoryear {%
{Takata}%
}{%
{Takata}%
}{%
{\protect \APACyear {2016}}%
}]{%
takata2016}
\APACinsertmetastar {%
takata2016}%
\begin{APACrefauthors}%
{Takata}, M.%
\end{APACrefauthors}%
\unskip\
\newblock
\APACrefYearMonthDay{2016}{{\APACmonth{12}}}{},
\newblock
\unskip
\newblock
\APACjournalVolNumPages{\pasj}{68}{6}{109}.
\newblock
\begin{APACrefDOI} \doi{10.1093/pasj/psw104} \end{APACrefDOI}
\PrintBackRefs{\CurrentBib}

\bibitem [\protect \citeauthoryear {%
{Tassoul}%
}{%
{Tassoul}%
}{%
{\protect \APACyear {1980}}%
}]{%
tassoul1980}
\APACinsertmetastar {%
tassoul1980}%
\begin{APACrefauthors}%
{Tassoul}, M.%
\end{APACrefauthors}%
\unskip\
\newblock
\APACrefYearMonthDay{1980}{{\APACmonth{08}}}{},
\newblock
\unskip
\newblock
\APACjournalVolNumPages{\apjs}{43}{}{469-490}.
\newblock
\begin{APACrefDOI} \doi{10.1086/190678} \end{APACrefDOI}
\PrintBackRefs{\CurrentBib}

\bibitem [\protect \citeauthoryear {%
{Unno}%
, {Osaki}%
, {Ando}%
, {Saio}%
\BCBL {}\ \BBA {} {Shibahashi}%
}{%
{Unno}%
\ \protect \BOthers {.}}{%
{\protect \APACyear {1989}}%
}]{%
unno1989}
\APACinsertmetastar {%
unno1989}%
\begin{APACrefauthors}%
{Unno}, W.%
, {Osaki}, Y.%
, {Ando}, H.%
, {Saio}, H.%
\BCBL {}\ \BBA {} {Shibahashi}, H.%
\end{APACrefauthors}%
\unskip\
\newblock
\APACrefYear{1989},
\newblock
\APACrefbtitle {{Nonradial oscillations of stars}} {{Nonradial oscillations of
  stars}}.
\newblock
\APACaddressPublisher{}{Tokyo: University of Tokyo Press, 1989, 2nd ed.}
\PrintBackRefs{\CurrentBib}

\end{thebibliography}

\end{document}